\documentclass[graybox]{svmult}

\usepackage{type1cm}        
\usepackage{makeidx}         
\usepackage{graphicx}        
\usepackage{multicol}        
\usepackage[bottom]{footmisc}

\usepackage{newtxtext}       %
\usepackage{newtxmath}       

\makeindex             

\newcommand{\boldr}{\mathbf {r}}
\newcommand{\bolds}{\mathbf {s}}

\newcommand{\boldv}{\mathbf {v}}
\newcommand{\boldx}{\mathbf {x}}
\newcommand{\boldR}{\mathbf {R}}
\newcommand{\boldt}{\mathbf {t}}
\newcommand{\rhat}{\hat{\mathbf {r}}}

\newcommand{\bom}{{\mbox{\boldmath $\omega$}}}

\newcommand{\boldS}{\mathbf {S}}

\begin{document}

\title*{TANGLED VORTEX LINES: DYNAMICS, GEOMETRY AND TOPOLOGY
OF QUANTUM TURBULENCE}
\titlerunning{Tangled quantum vortex lines} 
\author{Carlo F. Barenghi}

\authorrunning{Tangled vortex lines} 


\institute{Carlo F. Barenghi \at School of Mathematics, Statistics and
Physics and Joint Quantum Centre Durham-Newcastle,
Newcastle University, Newcastle-upon-Tyne, NE1 7RU, United Kingdom.
	\email{carlo.barenghi@newcastle.ac.uk}}

%
%
\maketitle


\abstract{ Near absolute zero, superfluid liquid helium displays
quantum properties at macroscopic length scales. One property, superfluidity,
means flow with zero viscosity. Another property, 
the existence of a complex wavefunction,
constrains the rotation to thin,
discrete vortex lines carrying one quantum of circulation each. Therefore,
if liquid helium is stirred, it becomes turbulent, but in a peculiar way:
it is a state of turbulence consisting of a tangle of quantised
vortex lines in a fluid without viscosity. Surprisingly, this disordered
state, which I called quantum turbulence years ago,
shares many properties with ordinary turbulence 
as it represents its essential skeleton.
These lectures attempt to relate the dynamics, the geometry and 
the topology of quantum turbulence. 
Although recently much progress has been made,
there are still many open questions.  Some of the methods which
have been used and are described here
(e.g. the smoothed vorticity, the use of crossing numbers) have
a scope which clearly goes beyond the fascinating problem of turbulence 
near absolute zero.
}

\section{Motivation: is turbulence knotted?}
\label{sec:motivation}

\begin{figure}[!ht]
    \begin{center}
    \scalebox{0.28}{\includegraphics{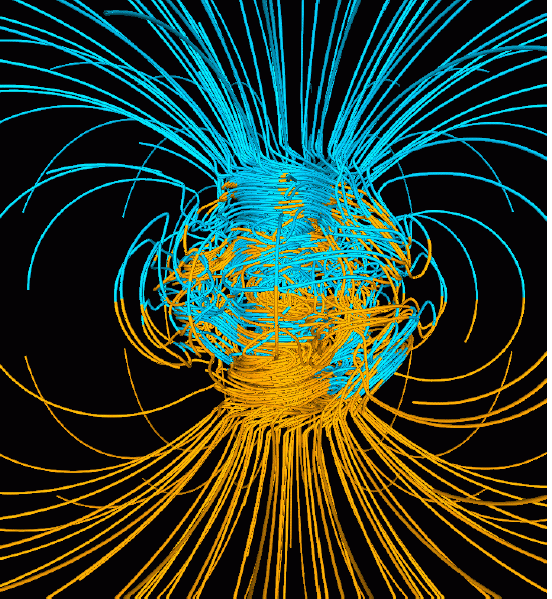}}
    \caption{Field lines of the Earth's magnetic field. The
    field lines are tangled in the Earth's core region where the turbulent 
    convection
    which generates the magnetic field takes place.
    The lines are colour-coded depending on the direction of the radial
    magnetic field.  From Ref.~\cite{GlatzmaierRoberts}.}
    \label{fig:GlatzmaierRoberts}
    \end{center}
\end{figure}

Images of turbulent flows display a high
degree of geometrical and topological complexity:
streamlines, vortex lines and magnetic
field lines are apparently chaotic, twisted, perhaps linked and knotted.
An example of such tangled field lines is shown 
in Fig.\ref{fig:GlatzmaierRoberts}
displaying the Earth's magnetic field in the core region. Are these
field lines knotted? If so, why?

Although the governing equations of motion are well-known,
the complicated geometries and topologies which are consequences of
the dynamics are often surprising.
The importance of turbulence in applications which range from weather 
forecasting
to astrophysics to engineering motivates the search for a better
understanding of the relation between dynamics, geometry and topology in
turbulent flows. A first step in
this direction is to find suitable ways to quantify
the complexity of field lines which are observed or numerically computed.

The aim of these lectures, which are dedicated to an audience
with a background in applied mathematics and classical physics,
is to shift the context slightly, from the
traditional setting of turbulent fluids which obey the Navier-Stokes
equation and the equations of magneto-hydrodynamics,
to the realm of turbulent {\it quantum fluids} where vortices are quantised.
We shall see that in these more exotic systems, the quantisation of the
circulation introduces a constrain which simplifies the task of
measuring the complexity of turbulence, at least in principle. The price
to pay is that we have to abandon the
familiar world of classical physics, and consider what happens to matter
at very low temperatures, a fascinating topic per se.

These lectures are organized in the following way. Section 2 contains a 
brief conceptual history of quantum fluids. Section 3 explains what is a 
quantum vortex (or vortex line) and compares it to an ordinary classical
vortex. Sections 4, 5 and 6 are dedicated respectively to the geometry,
the dynamics and the topology of turbulent tangles of quantum vortices.
Conclusions are (not) drawn in Section 7.

\section{A BRIEF HISTORY OF QUANTUM FLUIDS}
\label{sec:history}

The concept of absolute zero was probably invented by the French scientific
instrument maker Guillaume Amontons (1663-1705). He measured the changes
of pressure in a gas induced by changes of temperature, and argued that,
since the pressure cannot become negative, there must be a minimum
temperature, which he estimated at
$-240~\rm C$ (Celsius degrees). This value is
not far from the modern accepted value of absolute zero, which is
$-273.15~\rm C$, corresponding to $0~\rm K$ (Kelvin degrees).
The idea of absolute zero was consistent with the existence of atoms
and the kinetic theory of gases, pioneered by the Swiss mathematician
Daniel Bernoulli (1700-1782): if the temperature is the average kinetic energy
of the atoms, then, if motion ceases, temperature must vanish too.

Unfortunately the popularity
of the caloric theory of Antoine Lavoisier (1743-1794) halted the
development of the science of cold for a number of years.
Later, during the XIX century, progress restarted:
thermodynamics and then statistical mechanics convinced physicists that
if thermal disorder is reduced, the fundamental properties of matter
become more apparent.  This idea stimulated a race
to achieve lower and lower temperatures. Although the race was motivated by
fundamental physics, it indirectly
created technologies which changed life, e.g. the refrigeration of food.

Since a gas becomes a liquid if it is
cooled to a sufficiently low temperature, the liquefaction of the known gases
became milestones towards absolute zero. For example, in London, Michael Faraday
(1791-1867) liquefied ammonia at $-33~\rm C$ and
chlorine at $-34~\rm C$. In 1887,
Louis-Paul Cailletet (in Paris) and Raoul-Pierre Pictet (in Geneva)
liquefied oxygen at $-187~\rm C$. In 1877, in Krakow, Syzgunt von
Wroblewsky and Karol Olszewsky liquefied nitrogen at $-196~\rm C$.
The attention focused on the last remaining known gas to liquefy: hydrogen.
In 1898, in London, James Dewar liquefied hydrogen at $-253~\rm C$, and
thought he had won the race. However
the discovery of a new element, helium ($^4$He), first in the light from the
Sun then in Earth's minerals, soon restarted the race. Dewar's
main competitor was Heike Kamerlingh Onnes in Leiden. 
In 1908 Onnes
liquefied helium at $-270~\rm C$ and won the race. Soon after, in 1911,
Onnes also discovered that the electrical resistance of mercury vanishes
at temperatures lower than $4.2~\rm K$. He understood that he had found
a new state of matter, which he called a {\it superconductor}.

In the following years,
Onnes and collaborators realized that liquid helium acquires unusual
thermal and mechanical properties
below a critical temperature $T_{\lambda}=2.17~\rm K$ which they called 
the {\it lambda point} \footnote{The name comes from the shape of the 
specific heat vs temperature curve which has a sharp spike 
at $T=T_{\lambda}$.}
For example, if helium is cooled
by reducing the pressure of the vapour above the liquid
surface, as soon as the temperatures $T$ drops
below $T_{\lambda}$, the liquid surprisingly
stops boiling, becoming a perfect conductor of heat. Another example is
that, if further cooled, liquid helium does not become solid,
as all substances do.  To obtain solid helium, a great
pressure must be applied ($\approx 25~\rm atm$).
Physicists called {\it helium~II} the low
temperature phase of liquid helium below the lambda point, to distinguish
it from liquid helium above $T_{\lambda}$ which behaves like any ordinary
fluid and which they called {\it helium~I}. 
Hereafter we shall be concerned only
with helium~II.

On the theoretical side, in the early 1900s
the new science of quantum mechanics revealed that
matter has wave properties, for example a particle of mass $m$ and velocity
$v$ behaves like a wave of de Broglie wavelength $\lambda=h/(mv)$ where $h$ is
Planck's constant. According to statistical mechanics, the
average kinetic energy of a gas, $m v^2/2$, is proportional to $T$, hence
$\lambda \sim T^{-1/2}$. The natural question arises as to what happens as
$T \to 0$.  The question is particularly relevant for a gas of bosons
(particles with integer spin), which, unlike fermions (particles with
half-integer spin), are not prevented by Pauli's principle from
occupying the same quantum state, but can pile up into the ground state.
Satyendra Bose and Albert Einstein considered an ideal gas of boson particles.
They found that, at temperatures below a critical value $T_c$, the de Broglie
wavelength becomes larger then the average distance between the particles.
Essentially, the particles behave collectively like a giant
wave: the entire gas is governed by a macroscopic wavefunction $\Psi$,
creating a new state of matter which is now called
a {\it Bose-Einstein condensate} \cite{Pitaevskii,primer}.
As the temperature is reduced below $T_c$, a larger and larger fraction
of the particles condenses into the ground state; at $T=0$, all particles
are in the ground state.

In 1938 Peter Kapitza discovered that helium~II can flow through narrow
slits or thin capillaries as if it has zero viscosity. He called this property
{\it superfluidity} (motion without viscous dissipation), 
in analogy to {\it superconductivity} (motion without ohmic dissipation).
In the same year, Fritz London made
for the first time the link between helium~II and Bose-Einstein condensation.
He also highlighted the importance of the quantum mechanical zero point energy, 
which, in helium~II, is too large for the liquid to freeze and to
become a solid at normal pressure.

\begin{figure}[!ht]
    \begin{center}
    \scalebox{0.19}{\includegraphics{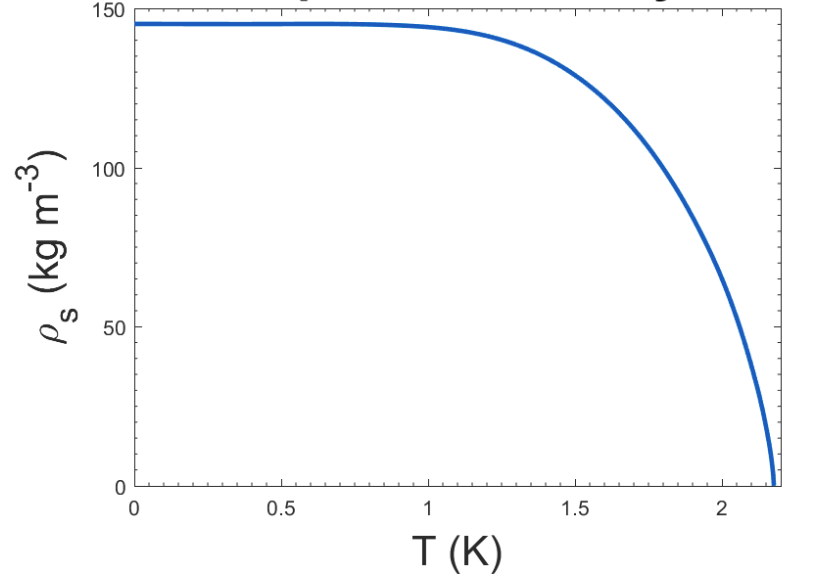}}
    \scalebox{0.19}{\includegraphics{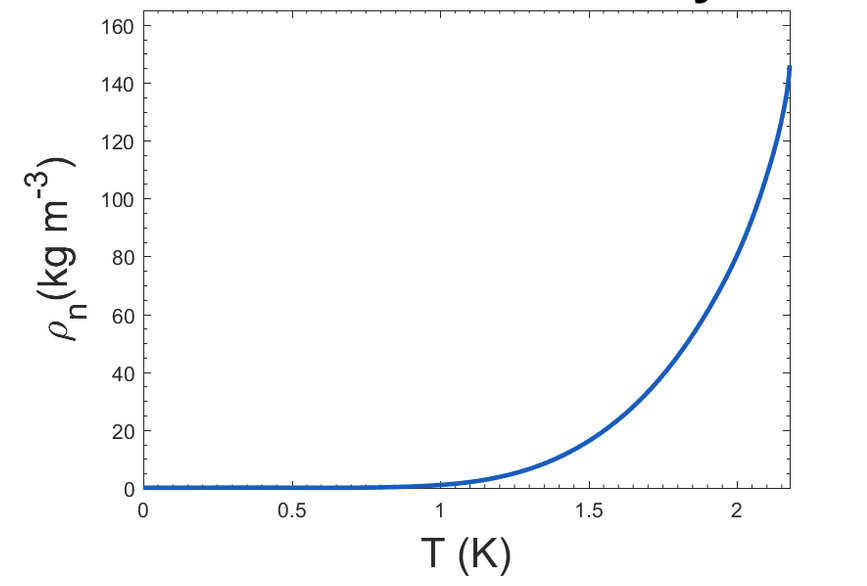}}
    \caption{Superfluid density $\rho_s$ (left) and normal fluid density
    $\rho_n$ (right) vs temperature $T$.}
    \label{fig:densities}
    \end{center}
\end{figure}

In the 1940s Lev Landau and Laszlo Tisza developed the {\it two-fluid model}
of helium~II, describing it as the mixture of two inter-penetrating fluid
components: the superfluid and the normal fluid. The model is
summarized by the following table:
\bigskip

\begin{center}
\begin{tabular}{| c | c | c | c | c | c |}
\hline
~fluid~component~   & ~density~  & ~velocity~   & ~viscosity~ & ~entropy~  \\  \hline
normal fluid & $\rho_n$ & $\boldv_n$ & $\eta_0$  & $S_0$     \\  \hline
superfluid   & $\rho_s$ & $\boldv_s$ & $0$       & $0$       \\ 
\hline
\end{tabular}
\end{center}
\bigskip

\noindent
Each fluid component has its own density and velocity fields: $\rho_s$,
$\boldv_s$ for the superfluid, and 
$\rho_n$, $\boldv_n$ for the normal fluid. The superfluid has zero entropy and
zero viscosity, whereas the normal fluid, consisting of thermal excitations,
is viscous and carries
the entire heat content of the liquid.
While the total density of liquid helium, $\rho=\rho_s+\rho_n$,
changes only a little with temperature, the
superfluid and normal fluid fractions, $\rho_s/\rho$ and $\rho_n/\rho$,
are rapid functions of
temperature: at $T=T_{\lambda}$ we have $\rho_s/\rho=0$ and $\rho_n/\rho=1$,
whereas at $T=0$ we have $\rho_s/\rho=1$ and $\rho_n/\rho=0$, as shown in
Fig.~\ref{fig:densities}. It is apparent from Fig.~\ref{fig:densities}
that for $T < 1~\rm K$ helium~II is effectively a pure superfluid. 
Physically, the inviscid superfluid component is related to the ground 
state\footnote{It must be noticed that 
the superfluid fraction is not the same as the 
condensate fraction: helium~II is a liquid, and the interaction
between atoms is much stronger than in a dilute gas.}
while the normal fluid consists of thermal excitations.
At small velocities, the superfluid can be modelled
by the classical Euler equation for an inviscid fluid, and the normal fluid
by the classical Navier-Stokes equation for a viscous fluid.

The two-fluid model explained the phenomenology
of helium~II (including for example the exceptional ability to transport heat,
which is exploited in many applications of cryogenic engineering)
and predicted new phenomena which were subsequently observed (for example,
second sound, an oscillation in which superfluid and normal fluid
move in antiphase, corresponding to a temperature wave at constant density
and pressure). 

Landau also understood that the property of superfluidity
arises from the dispersion relation of the elementary excitations. Using only
the laws of conservation of energy and momentum, he predicted that
an object moving slower than a certain critical velocity is unable to
reduce its energy by creating an excitation in the liquid, thus
perceiving the surrounding liquid as a vacuum.

In 1972 Douglas Osheroff, David Lee and Robert Richardson discovered that
the rare lighter isotope of helium, $^3$He, if cooled to temperatures
of few mK, also becomes superfluid. The effect is notable because $^3$He
is a fermion (the $^3$He nucleus contains two protons and only one neutron, 
not two), so Bose-Einstein condensation occurs via the formation of
Cooper pairs like in superconductors.

In 1995, Carl Wieman, Eric Cornell and Wolfgang Ketterle cooled
a very dilute weakly-interacting atomic gas to a temperature
of about 100 nK; for the first time they
experimentally achieved Bose-Einstein
condensation in a scenario similar to the ideal gas scenario which
was envisaged by Bose and Einstein.

In the following years other quantum fluids were discovered and
investigated. Besides $^4$He, $^3$He and a large number of
atomic condensates (lithium, sodium, potassium, rubidium, caesium, hydrogen,
etc), the realm of quantum fluids now comprises polaritons,
magnons, various two-component condensates, quantum ferrofluids, spinor
condensates, the
interior of neutron stars and probably even dark matter axions.

Hereafter, pursuing tangled field lines,
we shall focus the attention
on superfluid helium (i.e. helium~II, the low temperature
phase of liquid $^4$He), the system for which we have more experimental
and theoretical information about quantum vortices and turbulence. 
Occasionally we shall make reference to atomic condensates because
the main theoretical model of atomic condensates, 
the Gross-Pitaevski equation, is also used as a convenient qualitative model
of helium~II. For comparison, we shall also make reference to ordinary 
classical fluids obeying the Euler equation or the Navier-Stokes
equation.

\section{QUANTUM VORTICES}
\label{sec:vortices}

\subsection{Quantisation of the circulation}
\label{sub:quantisation}

Helium~II is remarkable not only for the absence of
viscosity, but also for the quantized vorticity. This property is
a consequence of the existence of a complex macroscopic wavefunction, $\Psi$,
which rules the system.
It is useful to write $\Psi$ in terms of its magnitude and phase as

\begin{equation}
\Psi(\boldx,t)=\vert \Psi(\boldx,t) \vert e^{i \phi(\boldx,t)},
\label{eq:psi}
\end{equation}

\noindent
where $\boldx$ and $t$ denote position and time respectively.
Because of the phase $\phi$, each part of the system "knows" what happens 
at other parts of the system (this is how a gas of individual atoms becomes
a condensate).

Using standard relations from quantum mechanics (Madelung transformation),
we define number density, $n(\boldx,t)$, and velocity, $\boldv(\boldx,t)$, as

\begin{equation}
n(\boldx,t)=\vert \Psi(\boldx,t) \vert^2,
\qquad
\boldv=\frac{\hbar}{m}\nabla \phi(\boldx,t),
\label{eq:Made}
\end{equation}

\noindent
where $\hbar=h/(2 \pi)$.
Since the curl of a gradient is always zero, Eq.~\ref{eq:Made}(b)
implies that the vorticity is zero. However, as we shall see,
this does not mean that the circulation is necessarily zero.
Consider a closed path C around
a simply-connected region S (i.e. a region S such that
any path C within S can be shrunk to a point which is inside S).
The circulation of the velocity around the path $C$ is
defined as

\begin{equation}
\Gamma=\oint_C \boldv \cdot {\bf dr}.
\end{equation}

\noindent
Since S is simply-connected, we can apply Stokes's Theorem,  use
$\nabla \times \boldv={\bf 0}$, and conclude that the circulation 
is zero:

\begin{equation}
\Gamma=\oint_C \boldv \cdot {\bf dr}=
\int_S (\nabla \times \boldv) \cdot d\boldS=0.
\end{equation}

\noindent
Now suppose that the region S is multiply connected,
i.e.  if we shrink C to a point, C may not remain within S.
In other words, suppose that
the region S contains a hole (a point or a region where $\Psi=0$,
hence, from Eq.~\ref{eq:Made}(a), the fluid's density is zero).
Then Stokes's Theorem cannot be applied; using Eq.~\ref{eq:Made}(b), 
we find that the circulation
is an integer multiple of a constant $\kappa=h/m$ called the
{\it quantum of circulation}:

\begin{equation}
     \Gamma=\oint_C \boldv \cdot {\bf dr}=
     \frac{\hbar}{m} \oint_C \nabla \phi \cdot {\bf dr}=
     \frac{\hbar}{m} \Delta \phi=
     \frac{\hbar}{m} 2 \pi q= q \kappa,
\end{equation}

\noindent
where $\Delta \phi=2 \pi q$ is the change of the phase going around 
the closed path C, and
$q$ is an integer (the last step follows from the fact that
the wavefunction is defined up to a phase factor which is either zero or
an integer multiple of $2 \pi$).
If $q$ is nonzero, according to  Eq.~\ref{eq:Made}(b),
the changing phase around the hole means
that there is a velocity field $v$ around the hole:
taking for $C$ a circle of radius $r$, we find that the azimuthal
velocity $v$ is

\begin{equation}
v=\frac{ q \kappa}{2 \pi r}.
\label{eq:velocity}
\end{equation}

\noindent
Eq.~\ref{eq:velocity} represents 
the well-known velocity field around a vortex line of 
fluid dynamics textbooks.  In most situations of interest, multi-charged
quantum vortices (quantum vortices with $q>1$) are energetically
unstable and decay in $q$ singly-charged ($q=1$) quantum vortices. Therefore
hereafter we assume that all quantum vortices are singly-charged ($q=1$).

\begin{figure}[!ht]
    \begin{center}
    \scalebox{0.24}{\includegraphics{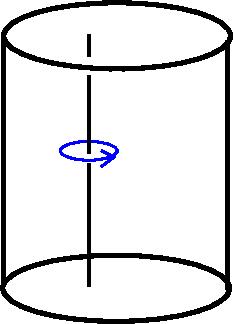}}~~
    \scalebox{0.24}{\includegraphics{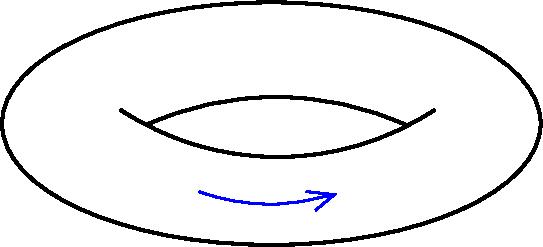}}
    \end{center}
    \caption{Examples of multiply-connected helium containers. The
     arrows indicate the direction of the flow corresponding to the changing
    phase around the hole. Left: a thin wire threads a cylindrical
    vessel from top to
    bottom\cite{Vinen1961,Hough2001}.
    Right: a torus-shaped vessel \cite{Schwab1997}.}
    \label{fig:multiply-connected}
\end{figure}

The next question is under which conditions a region of fluid can be
multiply-connected.
This happens if the hole is part of the experimental apparatus, as
shown in Fig.~\ref{fig:multiply-connected}.
At the left, the fluid is contained in a cylindrical vessel which is threaded
by a thin metal wire extending from the top to the bottom
\cite{Vinen1961,Hough2001}: clearly $\Psi=0$ in the region occupied
by the wire.
At the right, the fluid is inside a superfluid gyroscope, a device
shaped like a torus \cite{Schwab1997}.
In both cases, any flow around the wire or along the torus will have
quantised circulation.

\begin{figure}[!ht]
    \begin{center}
    \scalebox{0.22}{\includegraphics{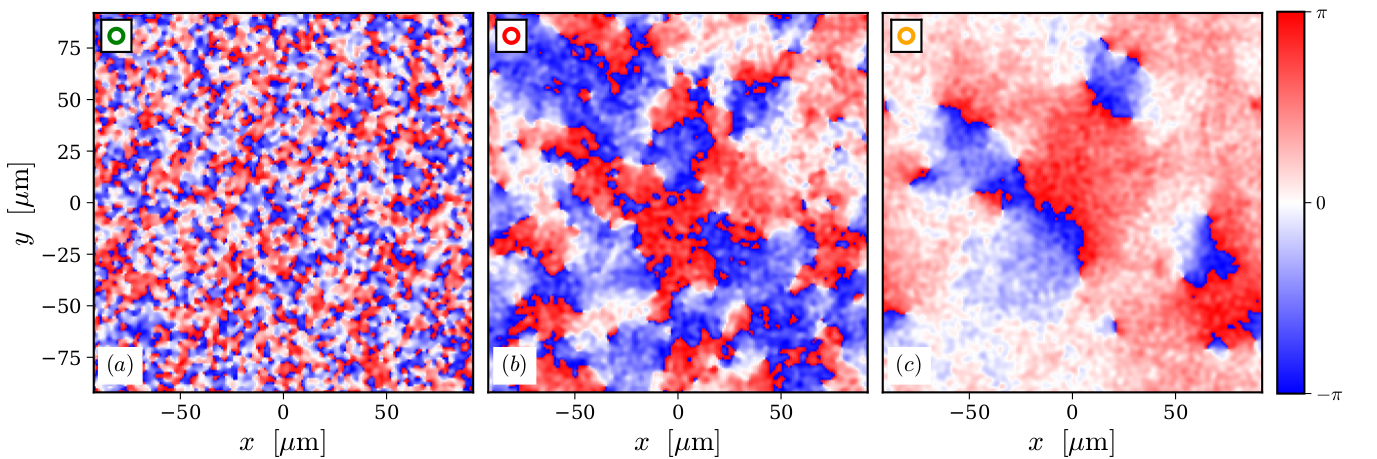}}
    \end{center}
    \caption{Phase distribution on the $xy$-plane as a gas of bosons
which is suddenly cooled through the critical temperature $T_c$. 
Initially (left) the phase is random, then the interactions reduce the
number of phase defects (middle) making them more visible (they
are the points around which the phase changes from $-\pi$ to $+\pi$);
at later stages (right) there are only few phase defects left.
From Ref.~\cite{Comaron2019}.}
    \label{fig:Comaron}
\end{figure}

A hole with a circulation around it can also form spontaneously in
the fluid: in this case we have a {\it quantum vortex}, 
also called a {\it vortex line}. 
The process which leads to the spontaneous formation
of a quantum vortex is called {\it vortex nucleation}. Vortex nucleation
occurs if an object inside the fluid moves at speed exceeding a certain
critical velocity \cite{Pomeau},
or, vice versa, if the fluid moves supercritically with
respect to the object or the boundary \cite{Keepfer}. Vortex nucleation
also occurs when the fluid is rapidly cooled across the critical temperature
$T_c$ \cite{Comaron2019}, as shown in Fig.~\ref{fig:Comaron}. 
Since the phase cannot simultaneously adjust to the same value
everywhere, phase domains will appear separated by  
points (in two-dimensions) or lines (in three-dimensions) 
where the wavefunction vanishes. Such points or lines are quantum vortices.
Since $\Psi=0$ implies that both the real part and the imaginary part
of $\Psi$ are zero, at these points the phase
$\phi=\tan^{-1}{( {\rm Im}(\Psi) / {\rm Re}(\Psi))}$
is not defined: this is why quantum vortices are called {\it phase defects}.

\begin{figure}[!ht]
    \begin{center}
    \scalebox{0.30}{\includegraphics{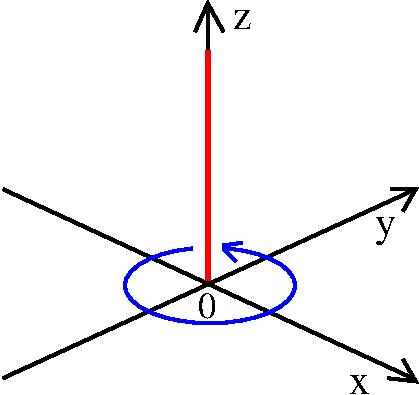}}\quad
    \scalebox{0.28}{\includegraphics{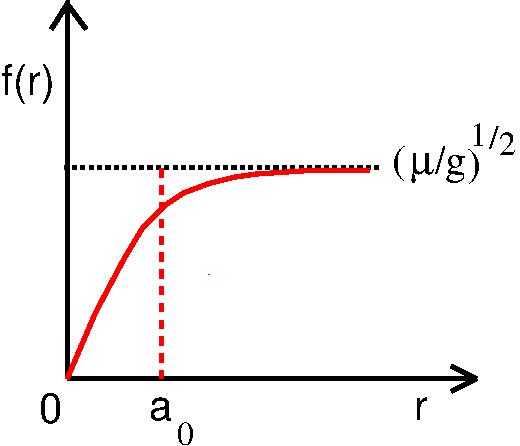}}
    \end{center}
    \caption{Left: vortex line aligned along the $z$-axis
(red line: the vortex core; blue line: the superfluid velocity).
Right: The solution $f\!(r)$ to Eq.~\ref{eq:vort} where $r=\sqrt{x^2+y^2}$
is the radial distance to the vortex axis, 
showing that the vortex core ($r<a_0$) is a region depleted of atoms.
On the vortex axis the density is exactly zero.}
    \label{fig:core}
\end{figure}

\subsection{The vortex core}
\label{sub:core}

To better understand the nature of a quantum vortex we use the
Gross-Pitaevskii equation (GPE), which is a model of
a weakly-interacting Bose gas at zero temperature \cite{primer}
(keeping in mind that although the GPE is a good quantitative model 
of atomic gases, it is only a qualitative model of helium~II):

\begin{equation}
i \hbar \frac{\partial \Psi}{\partial t}=
-\frac{\hbar^2}{2 m} \nabla^2 \Psi + g \vert \Psi \vert^2 \Psi -\mu \Psi.
\label{eq:GPE}
\end{equation}

\noindent
Here $m$ is the mass of one atom, $g=4 \pi \hbar^2 a_s/m$ is the
interaction parameter, $a_s>0$ is the
(repulsive) scattering length between the atoms,
and $\mu$ is the chemical potential. 
Using Eq.~\ref{eq:Made}, the GPE can be cast in the following
hydrodynamical form (this is why we use the term "quantum fluid"):

\begin{equation}
\frac{\partial \rho}{\partial t} + \nabla \cdot (\rho \boldv)=0,
\label{eq:cont}
\end{equation}

\begin{equation}
\frac{\partial \boldv}{\partial t}+ (\boldv \cdot \nabla)
\boldv =-\frac{1}{\rho} \nabla p + \frac{\hbar^2}{2m^2}
\nabla \left( \frac{\nabla^2 \sqrt{\rho}}{\sqrt{\rho}} \right),
\label{eq:modified-euler}
\end{equation}
\noindent
where $\rho=m n$ is the mass density, and the pressure is
$p=g \rho^2/(2 m^2)$.
Eq.~\ref{eq:cont} is the classical continuity equation
of fluid dynamics representing conservation of mass.
Without the second term on the right hand side (called the
{\it quantum pressure term}),
Eq.~\ref{eq:modified-euler} is the classical Euler
equation for a compressible inviscid fluid. It can be shown
that, at length scales larger than the {\it healing length}
$\xi=\hbar/\sqrt{g m n_0}=\hbar/\sqrt{\mu m}$, the quantum pressure
becomes negligible compared to the pressure, and we recover
classical Euler dynamics. Indeed, this is what we expect in
the limit as $\hbar \to 0$.
The quantum pressure is responsible for physical effects which
go beyond Euler dynamics, such as the structure of
the vortex core,
vortex nucleation and vortex reconnections.

The steady uniform solution of the GPE
is $\Psi_0=\sqrt{\mu/g}$, corresponding to
the number density $n_0=\vert \Psi_0 \vert^2=\mu/g$.
Using cylindrical coordinates,
the steady solution for a singly-charged ($q=1$) vortex placed
at the origin of a uniform condensate and aligned along the $z$-direction
as in  Fig.~\ref{fig:core}(left) is

\begin{equation}
\Psi(r,\theta)=f(r)e^{i \theta},
\end{equation}

\noindent
where $f(r)=\vert \Psi \vert$ satisfies

\begin{equation}
-\frac{\hbar^2}{2 m} \left( \frac{d^2f}{dr^2}+\frac{1}{r}\frac{df}{dr}
-\frac{f}{r^2} \right) + g f^3 -\mu f = 0.
\label{eq:vort}
\end{equation}

\noindent
The boundary conditions are that $f(r) \to 0$ for
$r \to 0$ and $f(r) \to \sqrt{\mu/g}$ for $r \to \infty$,
where $r$ is the distance to the origin.
By solving numerically Eq.~\ref{eq:vort}, we find that
$f(r)$ grows from $f=0$ at $r=0$ (the axis of the vortex)
to the bulk value $\sqrt{\mu/g}$ over a characteristic distance $a_0$
which is called the {\it vortex core radius}, see Fig.~\ref{fig:core}(right).
It turns out that the vortex core radius $a_0$ is of the order of the 
healing length $\xi$.

\begin{figure}[!ht]
    \begin{center}
    \scalebox{0.22}{\includegraphics{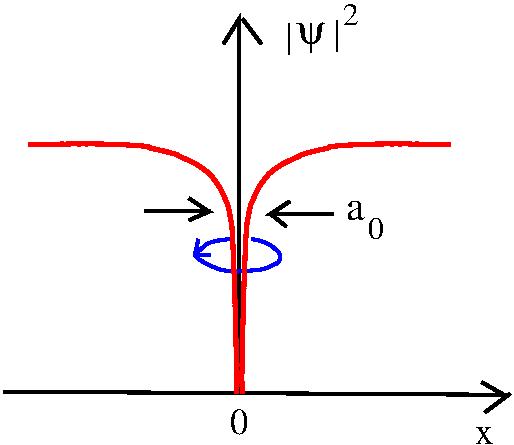}}
    \scalebox{0.19}{\includegraphics{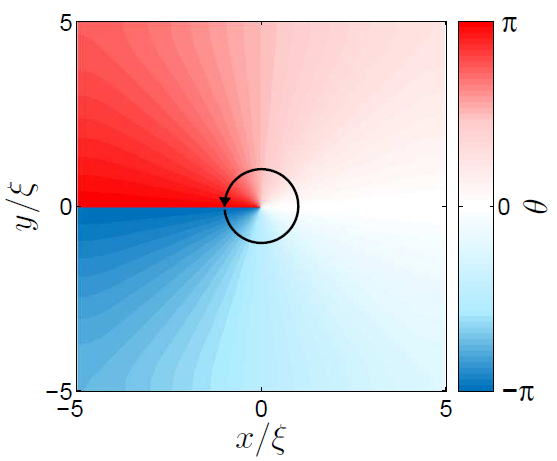}}
    \scalebox{0.22}{\includegraphics{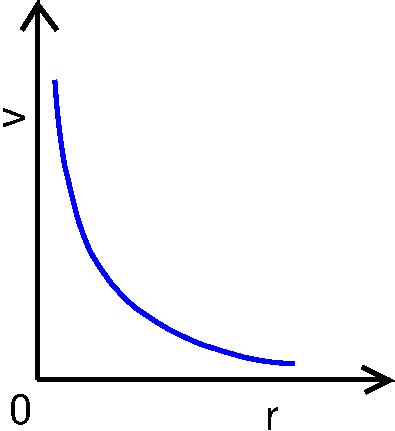}}\\
    \end{center}
    \caption{Vortex solution of the GPE. Left: density profile 
     $n=\vert \Psi \vert^2$ across the
     vortex.  Middle: phase on the $xy$-plane. Right: azimuthal velocity 
     $v=\kappa/(2 \pi r)$ vs $r$.}
\label{fig:vort}
\end{figure}

Fig.~\ref{fig:vort} schematically
describes the vortex solution of the GPE: a
density hole surrounded by a phase which changes from $0$ to
$2 \pi$, corresponding to the azimuthal velocity field
$v=\kappa/(2 \pi r)$. The fact that $v \to \infty$ as $r \to 0$ is not
a problem: since the density $n \to 0$ as
$r \to 0$, there are no atoms which move at infinite 
speed\footnote{The wavefunction $\Psi$ is well behaved on the vortex axis; 
what becomes singular 
when $r \to 0$ is the Madelung transformation, Eq.~\ref{eq:Made}.}.

\begin{figure}[!ht]
    \begin{center}
    \scalebox{0.22}{\includegraphics{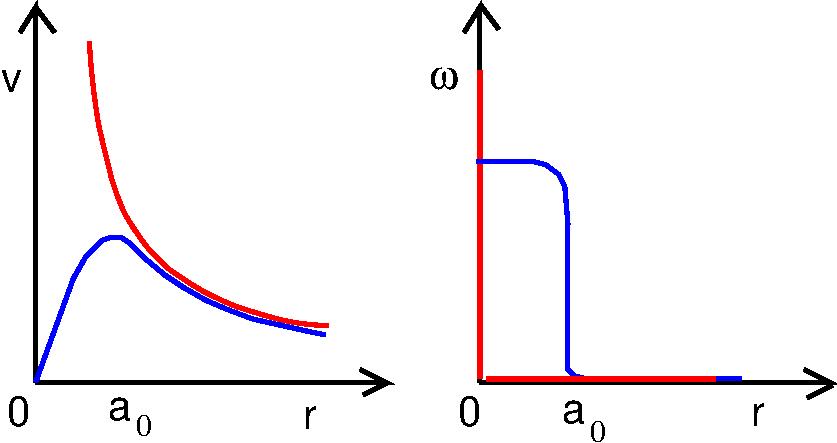}}
    \end{center}
    \caption{Azimuthal velocity $v$ (left) 
and axial vorticity $\bom$ (right) as a function
of radius $r$ corresponding to the quantum vortex solution
of the GPE (red), and the typical Rankine profile of classical fluids
(blue) which combines $v \sim r$ for $r < a_0$ and $v \sim 1/r$ for $r > a_0$. 
Note that the vorticity of the quantum vortex is zero for $r>0$
and nominally infinite at $r=0$.}
\label{fig:vort-om}
\end{figure}

It is instructive to keep track of the similarities and the differences
between quantum vortices and classical vortices (described by the
Euler equation or the Navier-Stokes equation).
Fig.~\ref{fig:vort-om} compares velocity and vorticity fields of a
classical vortex (blue curves) to a quantum
vortex solution of the GPE (red curves). 
The classical vortex combines an inner region
consisting of a solid-body rotating core (where the azimuthal velocity 
is proportional to $r$ and the axial vorticity is constant) with an outer
region which is irrotational (azimuthal velocity proportional to $1/r$ and
zero vorticity); the density can be constant at all values of $r$.
The quantum vortex has, by definition, density which vanishes at $r=0$, 
velocity which is strictly
proportional to $1/r$ everywhere, and 
delta-function vorticity concentrated
on the vortex axis at $r=0$.

In most experiments with atomic condensates, the size of the condensate,
$D$, is larger but not much larger than the vortex core size $a_0$;
typical values are $a_0\approx 0.1 \mu\rm m$ and $D$ ranging from
$D \approx 10 a_0$ to $\approx 100 a_0$; the average inter-vortex
distance, $\ell$, is typically $\ell \approx 10 a_0$.

In helium~II, the vortex core radius, 
$a_0 \approx 10^{-10}~\rm m$,
is many orders of magnitude
smaller than $D$ and $\ell$, typically
$D \approx 0.1$ to $10^{-3}~\rm m$
and $\ell \approx 10^{-3}$ to $10^{-5}~\rm m$. 
The smallness of $a_0$ compared to $D$ and $\ell$ is
the reason why the Vortex Filament Model
(see Section \ref{sec:dynamics}) is a good model of helium experiments.

\begin{figure}[!ht]
\scalebox{0.28}{\includegraphics{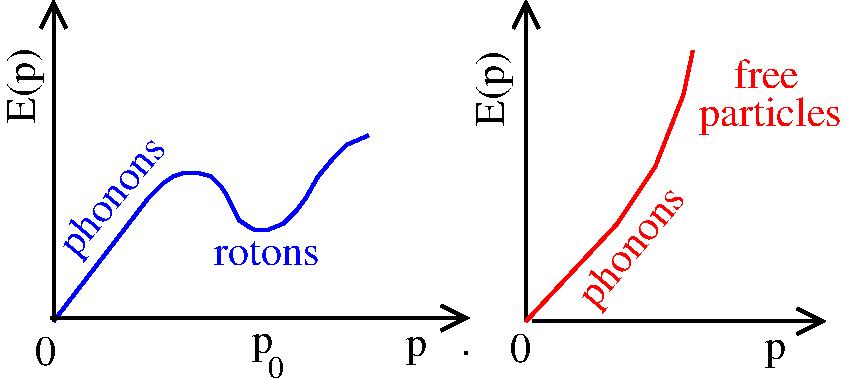}}
\caption{Left: dispersion relation of the elementary excitations in helium~II. 
Energy $E(p)$ vs momentum $p=\hbar k$ where $k$ is the
wavenumber. Note the phonon region, where $E \sim p$, and the roton's minimum
at  $p \approx p_0$.
Right: dispersion relation of the elementary excitations predicted by the
GPE. The phonon region is followed by free particles $E(p) \sim p^2$
without a roton minimum.}
\label{fig:spectrum}
\end{figure}

\subsection{More insight into the vortex core}
\label{sub:Reatto}

At this point it is worth exploring in more detail the nature of the
vortex core in helium~II beyond the simple description offered by the
GPE; this information will be useful when introducing helicity.
We have seen that the GPE predicts a depleted density profile with
zero density on the vortex axis, as in Fig.~\ref{fig:core} 
(right), and a velocity profile which diverges as $r \to 0$, as in
Fig.~\ref{fig:vort} (right).
It has already been mentioned that although the GPE
is a quantitative model of atomic gases, it is only a qualitative model
of helium~II: helium~II is a strongly-interacting liquid, not a dilute gas,
and the superfluid fraction is not exactly the condensate.   
The dispersion relation of the elementary excitations
of the uniform solution of the GPE consists of
phonons at small momentum $p$ and free particles at large $p$,
without the roton minimum which is characteristic of
liquid helium \cite{DonnellyDonnelly1981}, see Fig.~\ref{fig:spectrum}. 
Rotons require
a more sophisticated many-body quantum mechanical description 
which reveals that the vortex structure in helium~II
is more complex \cite{GalliReattoRossi2014,Amelio2018}
than predicted by the GPE.

\begin{figure}[!ht]
    \begin{center}
    \scalebox{0.28}{\includegraphics{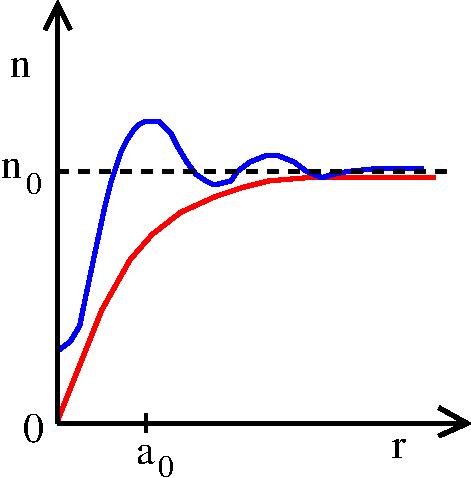}}
    \end{center}
\caption{Schematic profiles of the number density $n(r)$ vs radius $r$
near the vortex axis at $r=0$. The red line is the prediction of the GPE,
the blue line is the prediction of N-body quantum mechanics. Here
$n_0$ is the bulk number density and $a_0$ is the vortex core radius.}
\label{fig:vortex-Reatto}
\end{figure}

Consider a straight helium vortex.  The N-body wavefunction $\Psi(\boldR)$,
where $\boldR=(\boldr_1, \boldr_2, \cdots, \boldr_N)$ contains the coordinates
of $N$ atoms, must be an eigenstate of the angular momentum operator with
eigenvalues which are integer multiples of $N\hbar$.
Therefore $\Psi(\boldR)$ must be complex and have the form
$\Psi(\boldR)=\Psi_0(\boldR)e^{i \Omega(\boldR)}$, where $\Omega(\boldR)$
is the phase. The standard approach (called the fixed phase approximation)
consists of choosing
$\Omega(\boldR)$ and solving the resulting Schroedinger equation
for $\Psi_0(\boldR)$ allowing interatomic correlations at short
distances. The simplest choice for the phase is the Onsager-Feynman phase
$\Omega(\boldR)=\sum_{j=1}^N \theta_j$ where $\theta_j$ is the azimuthal angle
of atom $j$ with respect to the axis of the vortex. This choice
gives rise to a velocity field which is irrotational everywhere but on
the vortex axis, where it diverges, making the vorticity a delta function
localized on the vortex axis as in the GPE model.
The equation for $\Psi_0(\boldR)$ yields a density profile $n$ which vanishes
at the vortex axis (as in the GPE model), but which also displays
density oscillations near the edge of the core; these oscillations have
wavenumbers typical of rotons\footnote{It is interesting to notice that the
density oscillations and the roton
feature of the excitations are also captured by a variant of the GPE
called the Nonlocal Nonlinear Schroedinger Equation, an integral partial
differential equation which replaces
the hard-sphere collisions of the atoms modelled by the GPE
with a smoother Lennard-Jones
potential \cite{BerloffRoberts1999}).}.

This improved vortex model can be further refined 
(yielding lower vortex energy,
in better agreement with experiments) if the Onsager-Feynman assumption
is taken only as initial guess and inter-particle correlations are taken
into account also in determining the phase \cite{OrtizCeperley1995}.
This refined model yields features
\cite{OrtizCeperley1995,SaddChesterReatto1997} which
are schematically summarized in
Fig.~\ref{fig:vortex-Reatto}: the density in the
core is depleted (as in the GPE model) but remains nonzero on the vortex axis,
and the azimuthal velocity
acquires the classical
form of a Rankine vortex as in Fig.~\ref{fig:vort-om}, with
crossover from $v \sim r$ behaviour to $v \sim 1/r$ behaviour
at length scale $r \approx a_0$: this second feature means that the vorticity is
approximately constant inside the core. 

The conclusion is surprisingly simple: 
according to the best microscopic information available from N-body
quantum mechanics, the vortex
core in helium~II is similar to a classical vortex, unlike the GPE vortex
structure.

\section{GEOMETRY OF QUANTUM TURBULENCE}
\label{sec:geometry}

\subsection{In real space}
\label{sub:vortex-tangles}

It is relatively easy to excite turbulence in a condensate.
Helium~II can be made turbulent by applying a heat flow (as done
by W.F. Vinen, who pioneered the study of quantum turbulence),
or towing a grid, or rotating propellers,
or oscillating grids or forks, or applying ultrasound.
Atomic condensates can be made turbulent by stirring using
a laser spoon, or by oscillating or shaking the confining trap.
Experimental techniques exist to detect and visualize the quantum vortices
which are thus generated. Currently, the most popular visualization
techniques used in
helium~II are based on tracer particles (typically micron-sized
solid hydrogen particles which can become trapped in the vortex cores). 
Individual vortex lines can thus be visualized.
The local velocity can be measured using a variety of local
probes ranging from miniature Pitot tubes to cryogenic hot wires to 
mini cantilevers. Such measurements are crucial to determine the energy
spectrum, that is to say the distribution of the kinetic energy over
the length scales in a turbulent flow.

  In atomic condensates
absorption images are taken of the expanded condensate after the trap
has been switched off (the expansion is necessary to make the condensate
large enough to image it). Such visualization is necessarily destructive.
A non-destructive stroboscopic technique also exists which
can visualize a couple of vortices at the time. 
The main reason for which the study of quantum turbulence is more
advanced in helium~II than in atomic condensates is the lack
of three dimensional visualization and of local velocity probes in
atomic condensates. 

\begin{figure}[!ht]
    \begin{center}
    \scalebox{1.60}{\includegraphics{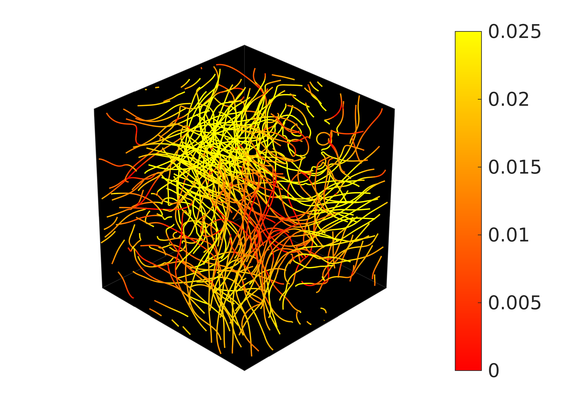}}
    \scalebox{0.90}{\includegraphics{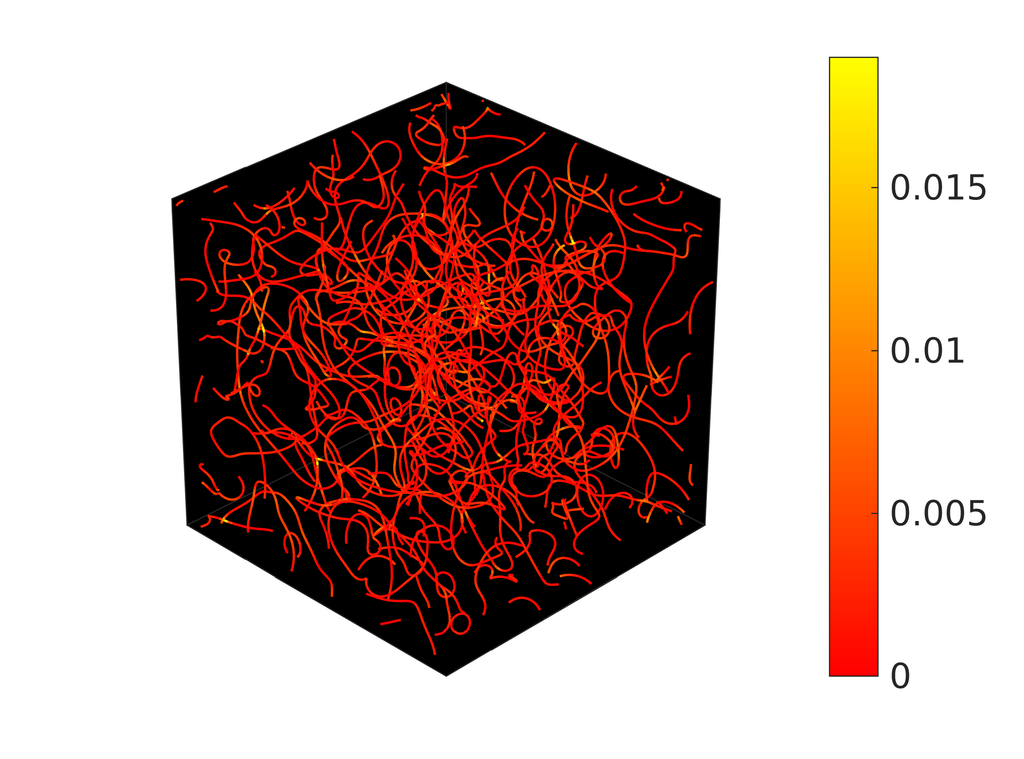}}\\
    \end{center}
    \caption{Snapshots of vortex tangles computed in a periodic cube
    using the Vortex Filament Model. The vortex lines are colour-coded
    according to the magnitude of the local helicity as explained in
    Section~\ref{sub:helicity}.}
\label{fig:tangles}
\end{figure}

In both helium~II and atomic condensates, turbulence manifests
itself as a disordered, time-dependent {\it tangle of vortex lines},
which move according to the velocity field which they generate and
reconnect when they collide. At nonzero temperatures the vortex
lines suffer a friction force when they move with respect to the
normal fluid \cite{friction}; 
microscopically, this friction is caused by the scattering
of phonons and rotons by the velocity field of the vortex lines.
Turbulence thus decays in time, unless it is continually
forced. Surprisingly, turbulence also decays at temperatures
below $1~\rm K$ where the normal fluid and the friction are negligible:
in Section~\ref{sub:sound} we shall see that moving vortices
generate sound waves (hence phonons, hence heat), thus the mechanism
that destroys the kinetic energy of the vortex lines is acoustic rather than viscous.

The results of numerical simulations of turbulence are
usually presented by displaying instantaneous snapshots of
vortex lines, see Fig.~\ref{fig:tangles}. These snapshots have been
generated using the Vortex Filament Model (see Section \ref{sec:dynamics}) 
using periodic boundary conditions
to represent a large homogeneous sample of helium~II:
lines which appear interrupted continue on the other side of the periodic
cube.  

To appreciate these images and the difference/similarities
between quantum turbulence and classical turbulence 
it is important to notice the following.
Firstly, the velocity field around each
vortex line thus displayed is constrained
to the irrotational form of Eq.~\ref{eq:velocity} (with $q=1$) (whereas
vortices in ordinary turbulence are unconstrained:
they can be large and small, weak and strong); secondly, the fluid which 
moves around each vortex line has zero viscosity (hence will not slow
down with time); 
thirdly, this visualization does not give information about the sense of 
rotation of each vortex line. 

\begin{figure}[!ht]
\centering
\begin{minipage}[t]{4cm}
\includegraphics[angle=0,width=0.9\textwidth]{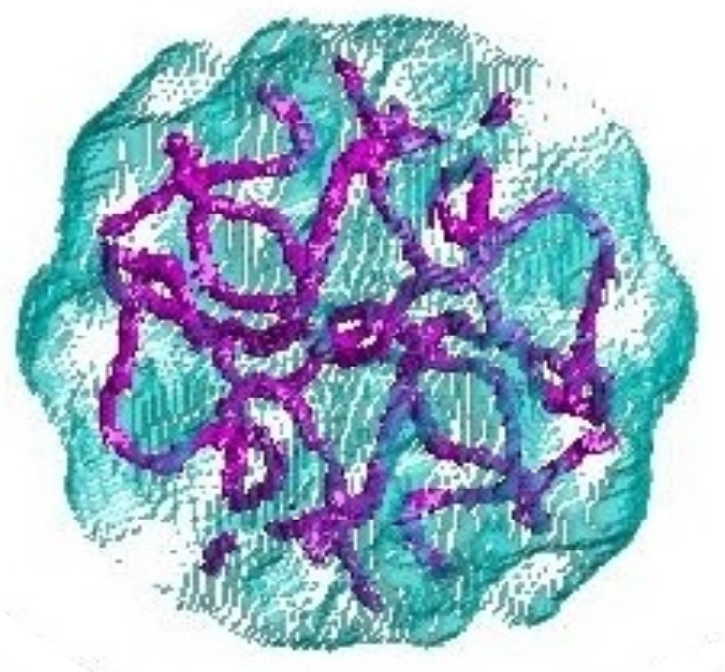}
\end{minipage}
\begin{minipage}[t]{4cm}
\includegraphics[angle=0,width=0.9\textwidth]{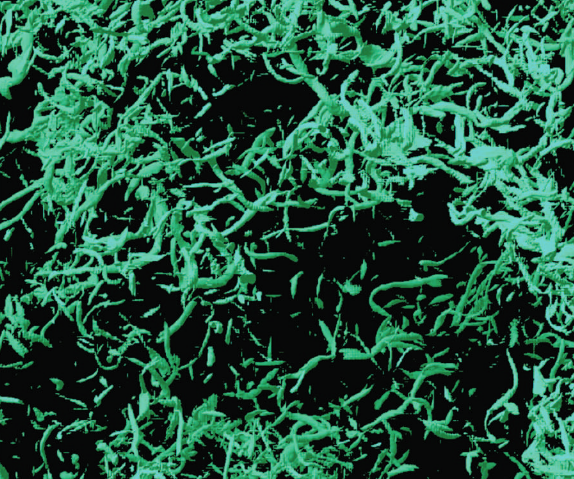}
\end{minipage}
\caption{Left: Snapshots of a vortex tangle in a spherically trapped
atomic condensate \cite{White2010}.  What is
plotted is the isosurface of the density (the outer surface of the
condensate in blue, the vortex cores in purple).
Right: Coherent vorticity structures in ordinary turbulence. 
Isosurfaces of the magnitude of the vorticity, computed at an
arbitrary threshold value, represent intense, localized tubular
regions of high rotation. From Ref.~\cite{Okamoto}.}
\label{fig:White-Okamoto}
\end{figure}

Fig.~\ref{fig:White-Okamoto} (left) shows turbulence in a spherically
trapped atomic condensate. 
What is displayed is the density isosurface representing the wobbly, 
oscillating outer part of the condensate (pale blue colour), and, 
inside the condensate, the cores of the vortex lines (purple colour). 
The quantum vortices appear as tubes, because the size of the condensate,
$D$, is only one order of magnitude larger than $a_0$.

It is interesting to compare quantum turbulence with ordinary turbulence
in terms of the geometry of the vorticity. As we have seen in
Fig.~\ref{fig:tangles} and \ref{fig:White-Okamoto} (left),
quantum turbulence consists of a tangle of vortex lines
constrained by the quantization of the circulation.
Ordinary turbulence, computed by solving the Navier-Stokes equation,
also contains tubular {\it coherent structures} or {\it vortex tubes}
 of intense vorticity, see
Fig.~\ref{fig:White-Okamoto} (right), somewhat similar to the vortex lines.
Since classical vorticity is continuous, classical turbulence
contains infinite vortex lines; thus these structures represent local 
bundles of parallel vortex lines, concentrating the vorticity enough
to be recognized by the plotting algorithm.
The size  of the coherent structures
is of the order of the Kolmogorov dissipation length scale \cite{Frisch}.
Although the number and shape and these structures depend on the arbitrary 
threshold value used for plotting, the analogy is intriguing.

In helium~II experiments and numerical simulations the intensity of the
turbulence is usually quantified by measuring the {\it vortex line density}
$L$, defined as the vortex length per unit volume. From $L$, one recovers
the {\it average inter-vortex distance}, $\ell \approx L^{-1/2}$. 
There is currently no experimental technique to
measure $L$ in atomic condensates, and this is one of the reasons for which
we focus on helium~II here instead.

\subsection{In k-space}
\label{sub:spectrum}

The apparently disordered
geometry of turbulence is better understood in Fourier space.
Indeed, the study of turbulence of the Navier-Stokes equation focuses on
the {\it energy spectrum}, $\hat E(k)$, which gives the distribution
of the kinetic energy over the length scales $2 \pi/k$ where
$k$ is the wavenumber. The prototype problem is homogeneous isotropic
turbulence, sustained in a statistical steady state
by injecting energy at some large length scale $D$  to compensate for
viscous losses. A brief review of classical turbulence
\cite{Frisch} is the following.  Let $E$ be the kinetic energy 
of the flow per unit mass.  After Fourier transforming $\boldv(\boldr,t)$,
the energy spectrum is defined by 

\begin{equation}
E=\frac{1}{V} \int_V \frac{1}{2} 
\boldv(\boldr,t) \cdot \boldv(\boldr,t) d^3\boldr
=\int_0^{\infty} {\hat E}(k) \, dk.
\end{equation}

\noindent
The energy which is injected at the scale $D$ is shifted by the
nonlinearities of the Navier-Stokes equation into smaller and smaller
scales, until, at some small length scale, $\eta$, called the 
{\it Kolmogorov length}, viscous forces dissipate the energy into heat.
This scale-by-scale transfer is called the {\it Richardson energy cascade}.
It can be shown that the energy spectrum ${\hat E}(k)$ satisfies the 
{\it Kolmogorov law}

\begin{equation}
{\hat E}(k)=C \epsilon^{2/3} k^{-5/3},
\label{eq:kolmo}
\end{equation}

\noindent
where $C$ is a dimensionless constant of order unity and $\epsilon=-dE/dt$
is the rate of energy dissipation (per unit mass).
It is also found that the ratio between the length scale of the energy input 
and the length
scale of the energy dissipation is $D/\eta \approx {\rm Re}^{3/4}$ where
${\rm Re}=UD/\nu$ is the {\it Reynolds number}, $U$ is the mean
flow, and $\nu$ the kinematic viscosity. 

\begin{figure}[!ht]
\centering
\includegraphics[angle=0,width=0.43\textwidth]{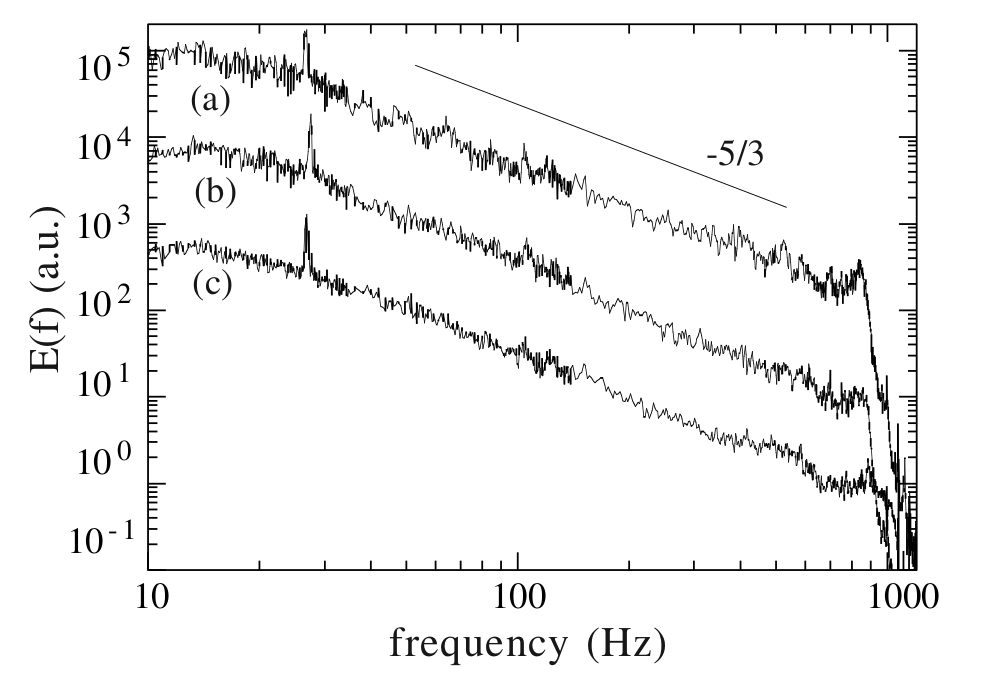}
\includegraphics[angle=0,width=0.4\textwidth]{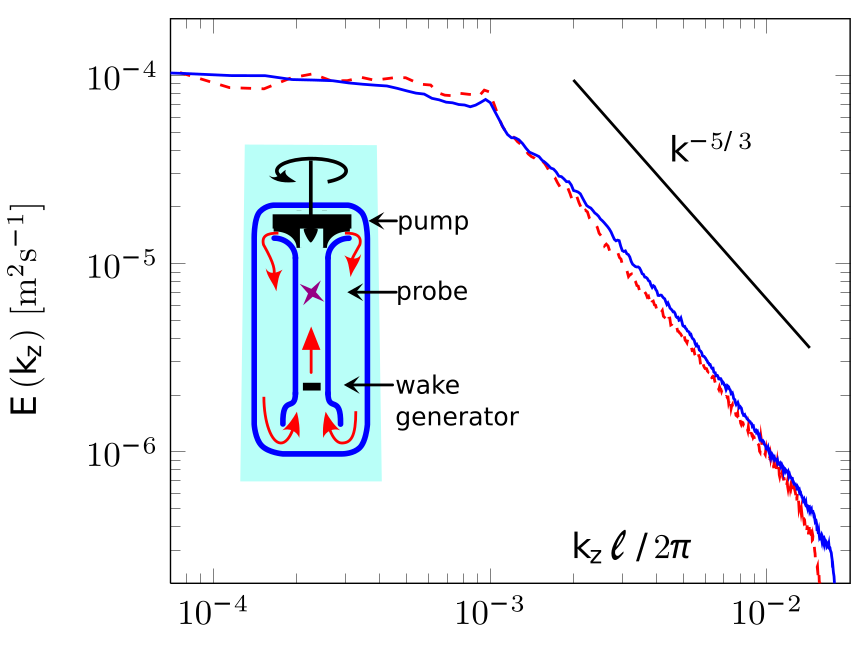}
\caption{Energy spectrum measured in helium stirred by counter-rotating
propellers at different temperature \cite{Tabeling} (left)
and in a superfluid wind tunnel \cite{Salort} (right).
In both cases, the spectrum displays a clear Kolmogorov
scaling ${\hat E}(k) \sim k^{-5/3}$.}
\label{fig:Tabeling-Salort}
\end{figure}

\begin{figure}[!ht]
\centering
\includegraphics[angle=0,width=0.36\textwidth]{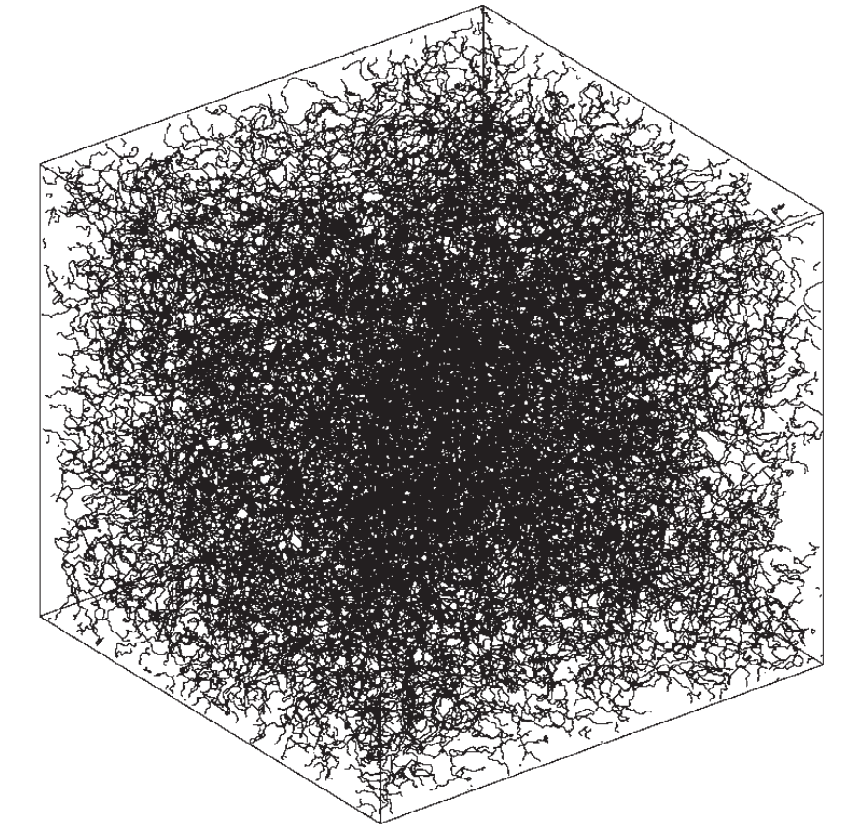}
\includegraphics[angle=0,width=0.4\textwidth]{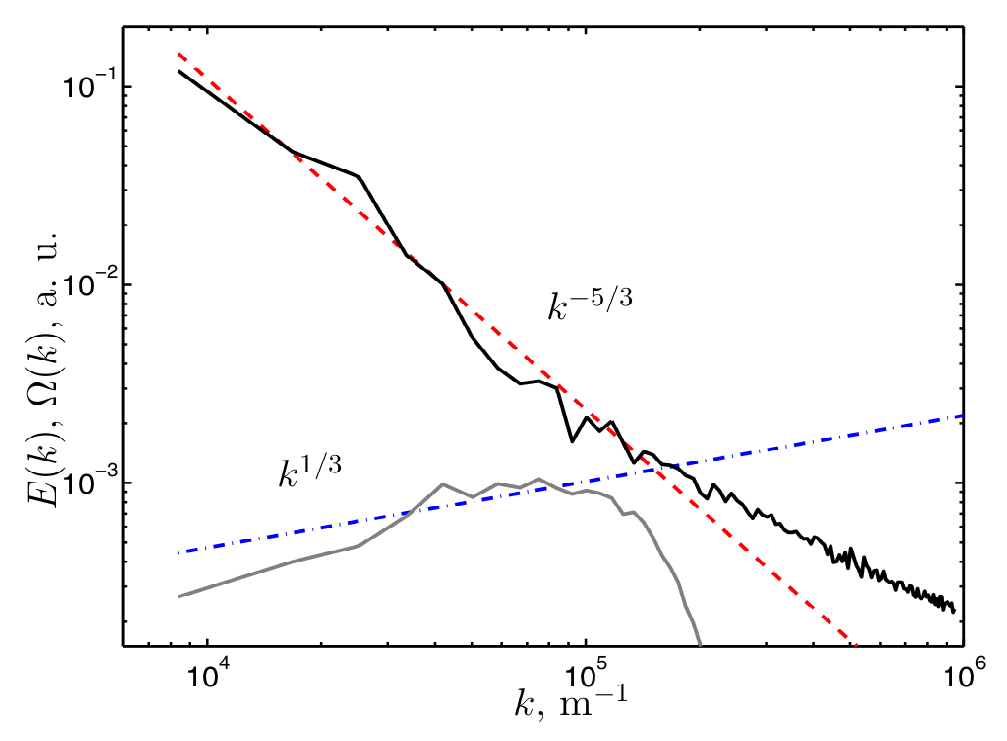}
\caption{Vortex tangle computed using the VFM
(left) displaying the classical Kolmogorov energy spectrum
${\hat E}(k) \sim k^{-5/3}$ (right) 
for $k<k_{\ell}=1.7 \times 10^5~\rm m^{-1}$ which crosses over to
${\hat E}(k) \sim k^{-1}$ for $k>k_{\ell}$. The curve below
is the enstrophy spectrum with the classical scaling
${\hat \Omega}(k) \sim k^{1/3}$ in $k_D <k<k_{\ell}$. The 
red and blue dashed lines
are guides to the eye to represent the $k^{-5/3}$ and the $k^{1/3}$
scalings respectively. From Ref.~\cite{Baggaley-EPL}.}
\label{fig:Baggaley-EPL}
\end{figure}

In helium, the Kolmogorov scaling has been observed in experiments 
\cite{Tabeling,Salort,Bradley}, see for example
Fig.~\ref{fig:Tabeling-Salort}. The scaling, which is considered the
signature of the energy cascade, has been confirmed by
numerical simulations \cite{Nore,Araki,Kobayashi} over the 
hydrodynamical range of length scales $k_D \ll k \ll k_{\ell}$,
where $k_D=2 \pi/D$ and $k_\ell=2 \pi/\ell$. At larger wavenumbers
$k>k_{\ell}$, the spectrum has the typical $k^{-1}$ behaviour of 
the spectrum of an individual
vortex line, as shown in Fig.~\ref{fig:Baggaley-EPL}. 

To get more insight into the geometry of the turbulence, in the
numerical simulations it is useful to take into account the 
orientation of the vortex lines by constructing the 
{\it smoothed vorticity} field, $\bom(\boldr,t)$ \cite{Baggaley-EPL}:

\begin{equation}
\bom(\boldr,t)=\kappa \sum_{j=1}^N
\frac{\bolds'_j}{(2 \pi \sigma^2)^{3/2} }
e^{-\vert \bolds_j-\boldr   \vert^2/(2 \sigma^2)}\, \Delta \zeta.
\end{equation}

\noindent
The notation used in this equation will be explained in Section~\ref{sub:vfm};
here it suffices to say that $\bolds_j$ ($j=1, \cdots, N$) is a
discretization point along the vortex lines, and $\bolds'_j$ is the
tangent unit vector to the line at $\bolds_j$.
Essentially, each vortex line is dressed
with an oriented Gaussian function which "smooths" it
over the length scale $\sigma \approx \ell$. 
Two parallel vortex lines  
create a strong smoothed vorticity $\bom$, while two antiparallel lines
give a negligible $\bom$.  From $\bom$, one defines
the {\it enstrophy} $\Omega$ and the enstrophy spectrum
${\hat \Omega}(k)$:

\begin{equation}
\Omega=\frac{1}{V} \int \bom(\boldr,t) \cdot \bom(\boldr,t) \, d^3 \boldr=
\int_0^{\infty} {\hat \Omega}(k) \, dk.
\end{equation}

\noindent
It is found \cite{Baggaley-EPL} that the enstrophy spectrum
has indeed the classical ${\hat \Omega}(k) \sim k^{1/3}$ scaling
which is expected from the $k^{-5/3}$ Kolmogorov spectrum of the energy
in the same range of $k$, see the bottom curve in Fig.~\ref{fig:Baggaley-EPL}.

\begin{figure}[!ht]
\centering
\includegraphics[angle=0,width=0.4\textwidth]{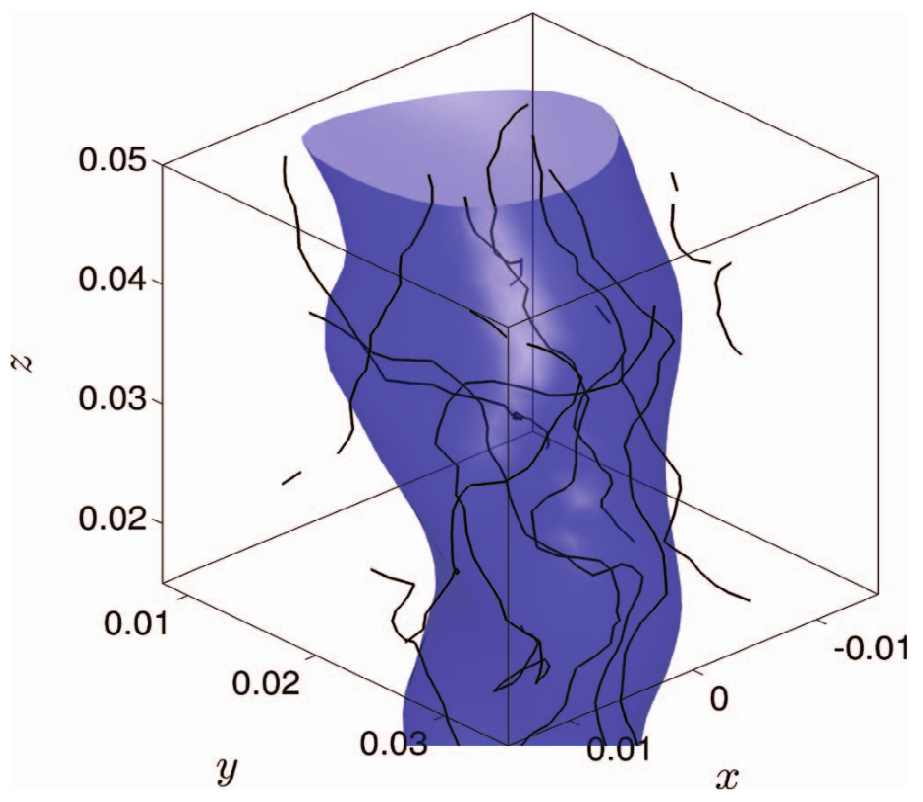}~~~~~~~~~
\includegraphics[angle=0,width=0.4\textwidth]{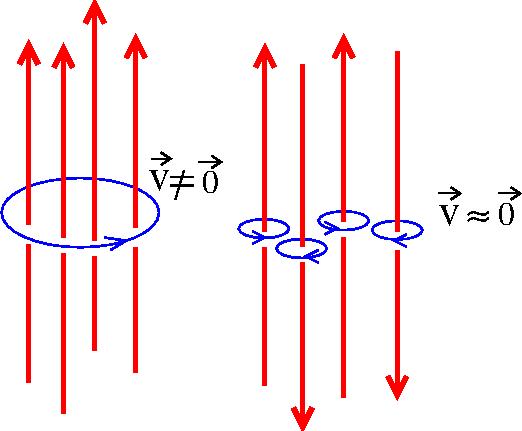}
\caption{Left: magnified plot of a vortex bundle in a numerical simulation
of turbulence performed using the VFM. The bundle is within the coloured
region, which is an isosurface of the smoothed vorticity. Only the
vortex strands which are within this region are plotted, i.e. these
vortex strands are parallel. From Ref.~\cite{Baggaley2014}.
Right: schematic bundles of vortex lines (the arrows show the direction
of the vorticity). If the vortices are parallel,
a large scale flow around them is created; if the orientation of the
lines is random, the velocity contributions cancel out.}
\label{fig:bundle}
\end{figure}

The conclusion \cite{BarenghiLvovRoche}
from these and similar experimental and numerical
studies is that helium~II,
at length scales larger than the average inter-vortex distance $\ell$,
behaves like a classical Navier-Stokes fluid displaying the Kolmogorov
$k^{-5/3}$ scaling (at shorter length scales, for $k>k_{\ell}$,
the dynamics of individual vortex lines takes over). The question is:
why?

The answer is the spontaneous partial polarisation of
the vortex lines. When strong, this polarization may even take the form of
visible bundles of parallel vortices.
The bundles may last only a brief time
before disappearing, but form again somewhere else in the flow. An
example of a vortex from numerical simulations
is shown in Fig.~\ref{fig:bundle} (left).
Within the bundle, the vortex lines are parallel, creating a relative
large flow as explained in Fig.~\ref{fig:bundle} (right). The bundles
contain energy.

\begin{figure}[!ht]
\centering
\includegraphics[angle=0,width=0.7\textwidth]{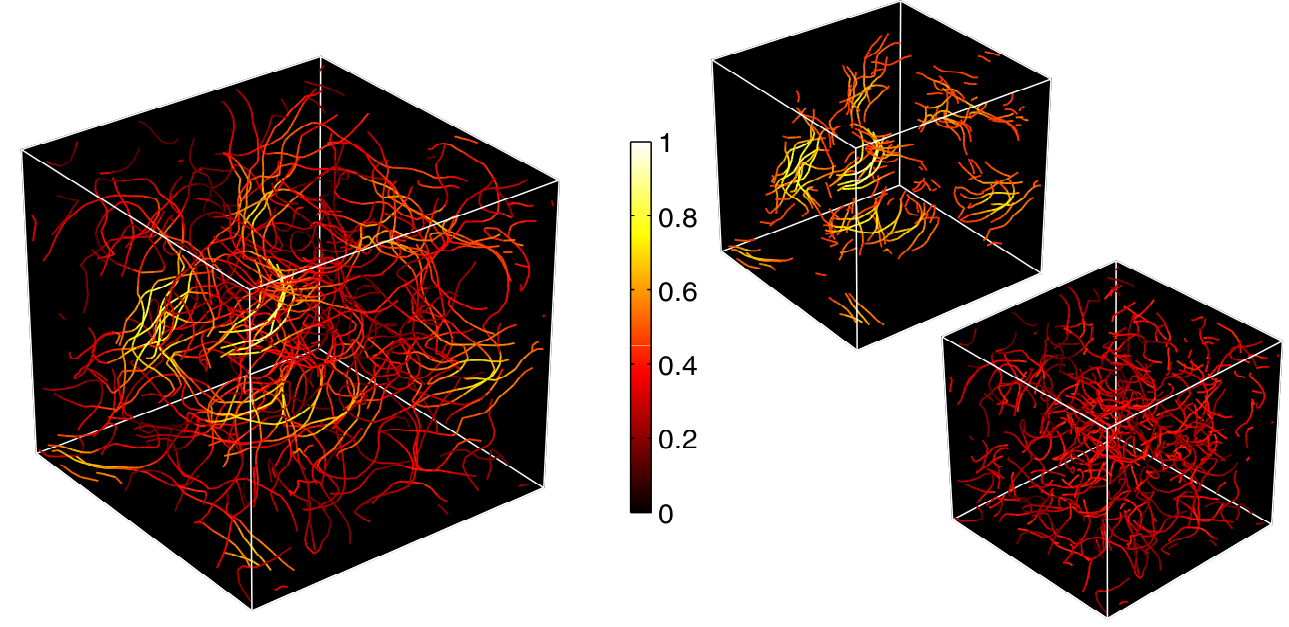}
\caption{Left: snapshot of a vortex tangle computed using the VFM. 
The vortex lines are
colour-coded according to the local smoothed vorticity (yellow regions
correspond to large values, red regions to low values).
On the middle and right, the same snapshot is split into locally
polarised lines ($\bom(\bolds_j,t) > 1.4~\bom_{rms}$) (middle) and random
lines ($\bom(\bolds_j,t) < 1.4~\bom_{rms}$) (right). The entire
tangle (left) and the polarised lines (middle) have a $k^{-5/3}$
energy spectrum, while the random lines (right) have a $k^{-1}$
spectrum. From \cite{Laurie}.}
\label{fig:Laurie}
\end{figure}

In a follow-up numerical
investigation \cite{Laurie}, a vortex tangle with energy spectrum
satisfying the Kolmogorov law was analysed
by instantaneously dividing the vortex lines in two groups:
polarised and unpolarised (a distinction  made by thresholding
the smoothed vorticity), as  shown in Fig.~\ref{fig:Laurie}. 
It was found that, when considered
separately, the polarised lines have a $k^{-5/3}$ spectrum, while the 
unpolarised lines have a $k^{-1}$ spectrum. This suggests that
the Kolmogorov spectrum, which piles up energy at small $k$,
is the effect of some partial polarization
of the vortex tangle (indeed, experimental
data \cite{Roche} suggests that most of the energy is carried by the
polarised lines). In a disordered vortex configuration, instead,
the $k^{-1}$ spectrum is a signature
of randomness: it means that the velocity field at a point is dominated
by the nearest vortex lines (the contribution of all other
lines cancel out). 

The classical behaviour of turbulent helium~II
at length scales $k_D < k < k_{\ell}$ is confirmed by studies of the velocity
statistics. The initial studies were puzzling: it was found that
the normalised histograms of the velocity components display
non-classical power law behaviour in the experiments \cite{Paoletti}
and in the numerical simulations \cite{White2010}, unlike the Gaussian
behaviour observed in classical turbulence \cite{Vincent}. But the explanation
was soon found:
in the experiments the velocity components are measured
used tracer particles which are smaller than the inter-vortex distance $\ell$,
and in the simulations they are naturally computed at selected
points in space. If the velocity components
are averaged over length scales larger than $\ell$ \cite{stats,LaMantia},
classical Gaussian statistics are recovered. It is therefore important
to keep in mind the distinction \cite{Skrbek2021} between
{\it classical length scales} ($k<k_{\ell}$) and {\it quantum length scales}
($k>k_{\ell}$).

\begin{figure}[!ht]
\centering
\includegraphics[angle=0,width=0.4\textwidth]{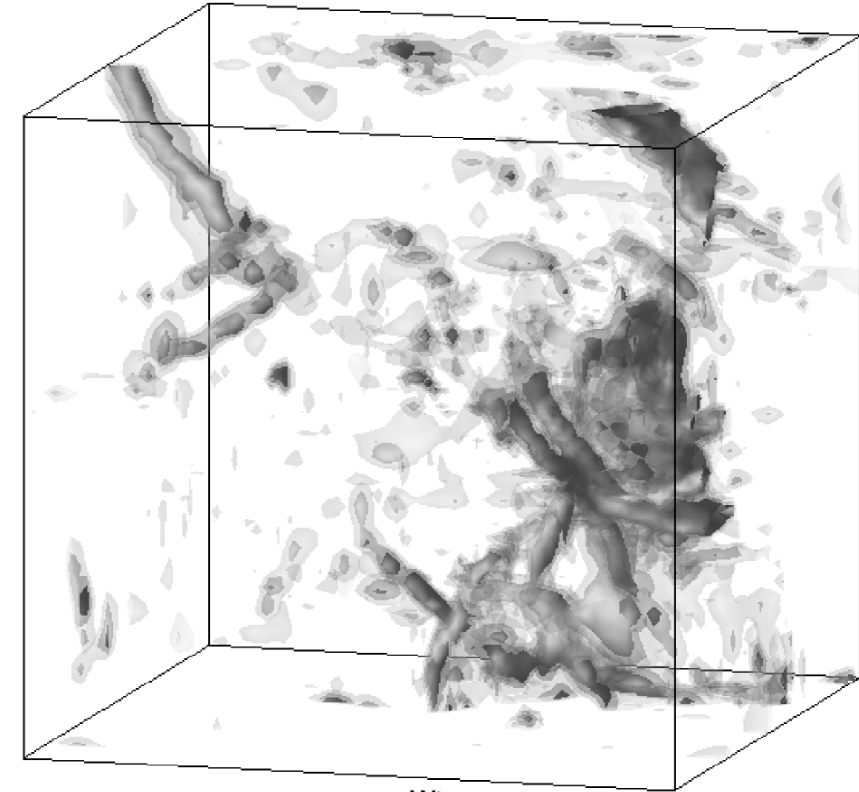}
\caption{Snapshot of the coherent part of classical turbulence
consisting of vortex tubes, responsible for the observed Kolmogorov
$k^{-5/3}$ energy spectrum. From \cite{Farge}.}
\label{fig:Farge}
\end{figure}

These results are consistent with ordinary turbulence.
Farge et al. \cite{Farge} used wavelets to decompose
classical Navier-Stokes turbulence
into a coherent part (see Fig.~\ref{fig:Farge})
and an incoherent part; it was found that the coherent part, consisting
of "vortex tubes", is responsible for most of the energy of the flow 
and for the observed Kolmogorov spectrum of the entire flow.

\subsection{Vinen turbulence}
\label{sub:vinen}

We should not rush to the conclusion that quantum turbulence is quite similar
to classical turbulence, at least for $k<k_{\ell}$. A second
regime of quantum turbulence has been
identified, which we shall refer to as {\it Vinen turbulence} to
distinguish it from the regime described in the previous section
which we shall refer to as {\it Kolmogorov turbulence}.  
Vinen turbulence has been observed in experiments with vortex rings
\cite{Walmsley}, experiments in $^3$He \cite{Bradley}, 
numerical simulations based on the VFM \cite{BaggaleyPRB2012},
GPE simulations of the thermal quench of a Bose gas \cite{Stagg},
GPE simulations of turbulence in an atomic
condensate \cite{Cidrim}, and in simulations of dark matter \cite{Mocz}.
The Vinen and Kolmogorov regimes
can be experimentally distinguished  by the rate of decay of the
vortex line density, which is either $L \sim t^{-1}$ (Kolmogorov)
or $L \sim t^{-3/2}$ (Vinen).

\begin{figure}[!ht]
\centering
\includegraphics[angle=0,width=0.4\textwidth]{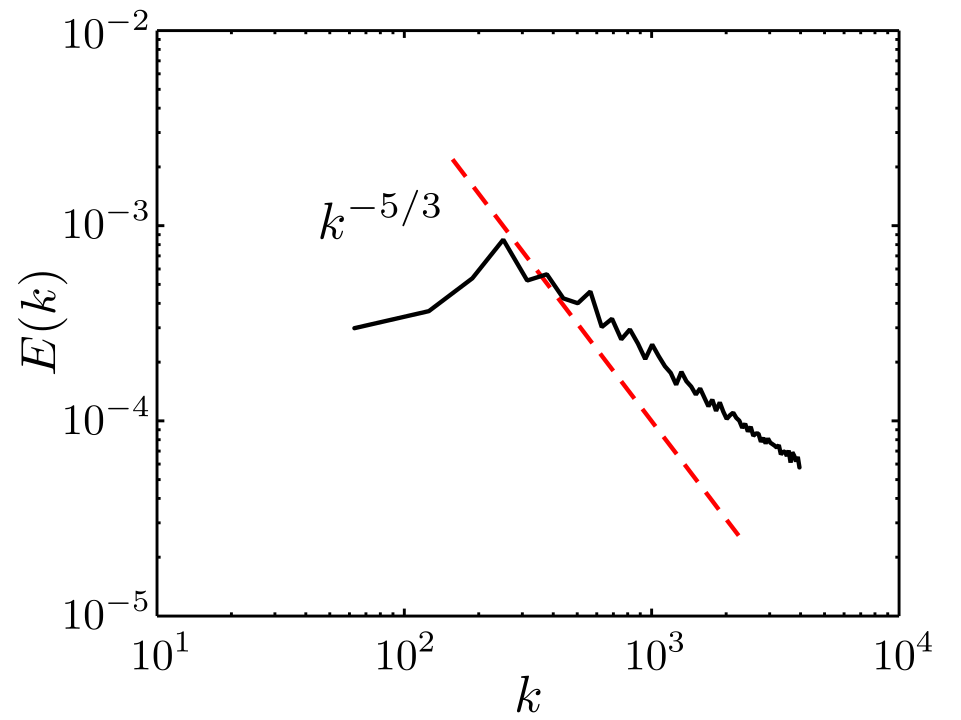}
\caption{Energy spectrum of Vinen turbulence. The dashed line is a guid to the
eye to indicate the Kolmogorov $k^{-5/3}$ scaling.
Notice the lack of energy
at the large length scales (small $k$), and the $k^{-1}$ behaviour at
large $k$.}
\label{fig:Vinen-spectrum}
\end{figure}

Vinen turbulence is different from Kolmogorov turbulence because it lacks
a clear energy cascade \cite{nocascade}. Typically, this happens
because there is no forcing at a large enough length scale, or
k-space is too limited. The spectrum of Vinen turbulence thus
lacks energy at the large length scales, peaks at intermediate
length scales around $k_{\ell}$, and decays as $k^{-1}$
for $k>k_{\ell}$, as shown in Fig.~\ref{fig:Vinen-spectrum}. 
Numerical simulations \cite{Stagg,Cidrim}
show that velocity correlations decay rapidly at distances larger
than $\ell$.

\begin{figure}[!ht]
\centering
\includegraphics[angle=0,width=0.4\textwidth]{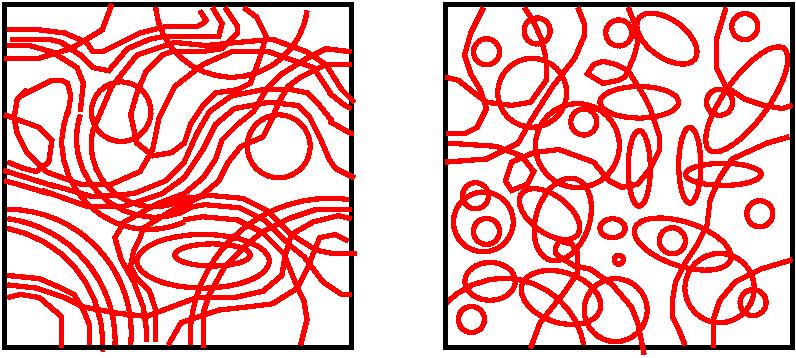}
\caption{Artistic representation of vortex lines in Kolmogorov turbulence (left)
and Vinen turbulence (right).}
\label{fig:volovik}
\end{figure}

It is probably correct to think of Vinen turbulence as consisting
more of random vortex rings than polarised lines or bundles, as is
shown schematically in Fig.~\ref{fig:volovik}.

\section{DYNAMICS OF QUANTUM TURBULENCE}
\label{sec:dynamics}

The dynamics of quantum vortices in a turbulent tangle is rich
in interesting effects. We have seen in Section~\ref{sec:geometry}
that a vortex tangle consists of a mixture of
interacting vortex lines and vortex loops. It is instructive to
consider the elementary physical processes which 
continually take place within the vortex tangle.

\subsection{Friction}
\label{sub:friction}

\begin{figure}[!ht]
\centering
\includegraphics[angle=0,width=0.35\textwidth]{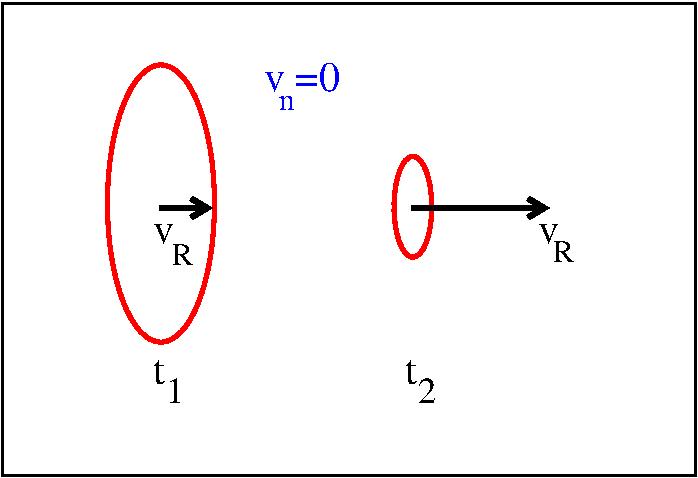}~~~~~~~
\includegraphics[angle=0,width=0.35\textwidth]{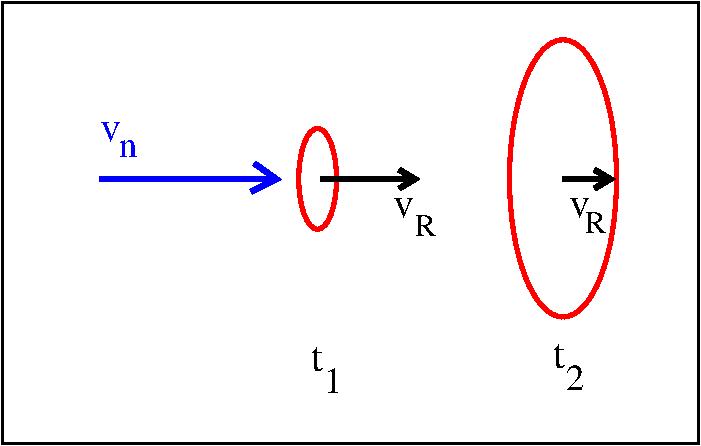}
\caption{Left: schematic vortex ring shown at two successive times 
$t_1<t_2$ travelling
from left to right as it decays due to friction with the stationary
normal fluid ($v_n=0$).
The black arrow is the ring's translational velocity $v_R$,
which increases as the ring's radius $R$ decreases.
Right: schematic vortex ring shown at two successive times 
$t_1<t_2$ travelling
from left to right as it gains energy from the normal fluid moving 
at velocity $v_n>v_R$ (blue arrow) in the same direction.
The black arrow is the ring's translational velocity $v_R$,
which decreases as the radius $R$ increases.}
\label{fig:ring}
\end{figure}

At sufficiently low temperatures 
the friction between the normal fluid and the vortex
lines tends to zero. In this temperature regime,
vortex rings of radius $R$, having energy proportional to $R$,  move
at translational self-induced velocity $v_R \sim 1/R$ (the larger the radius,
the slower the velocity).
More in general, we shall see in Section~\ref{sub:vfm} that, 
in the first approximation, curved vortex lines move in the binormal 
direction at velocity which is inversally proportional to the local radius 
of curvature. Therefore, when looking at images of vortex tangles as in
Fig.~\ref{fig:tangles}, we should recognize that
gently curved lines and large  loops move slowly, while
wiggly lines and small loops move rapidly. 
At higher temperatures, vortex lines which move with respect to the
stationary normal fluid lose energy to it. For example, a vortex ring
shrinks in size and speeds up, until it vanishes,
as is shown in Fig.~\ref{fig:ring} (left). Similarly, the amplitude of Kelvin
waves (see Section~\ref{sub:Kelvin}) decreases,
straightening vortices.

However, the friction between the superfluid and
the normal fluid can also transfer energy in the opposite direction: 
a moving normal fluid can feed energy
into the vortex tangle. For example, if the normal fluid velocity, $v_n$,
exceeds the self-induced velocity of the vortex ring, $v_R$, the ring
gains energy and its radius grows \cite{friction}, as is shown in
Fig.~\ref{fig:ring} (right).
Similarly, if the normal fluid velocity in the direction along a
vortex line is large enough, infinitesimal Kelvin waves along the line become
unstable and grow in amplitude ({\it Donnelly-Glaberson instability}).

\subsection{Kelvin waves}
\label{sub:Kelvin}

\begin{figure}[!ht]
\centering
\includegraphics[angle=0,width=0.20\textwidth]{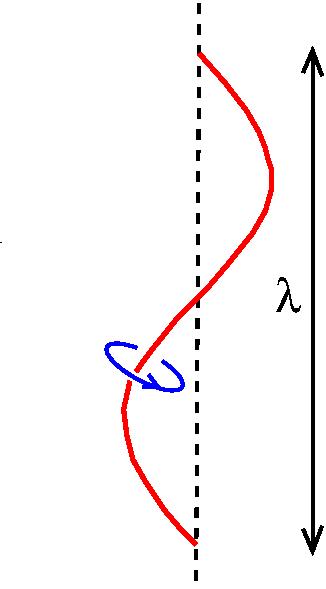}~~~~~~~~~~~~~~~
\includegraphics[angle=0,width=0.32\textwidth]{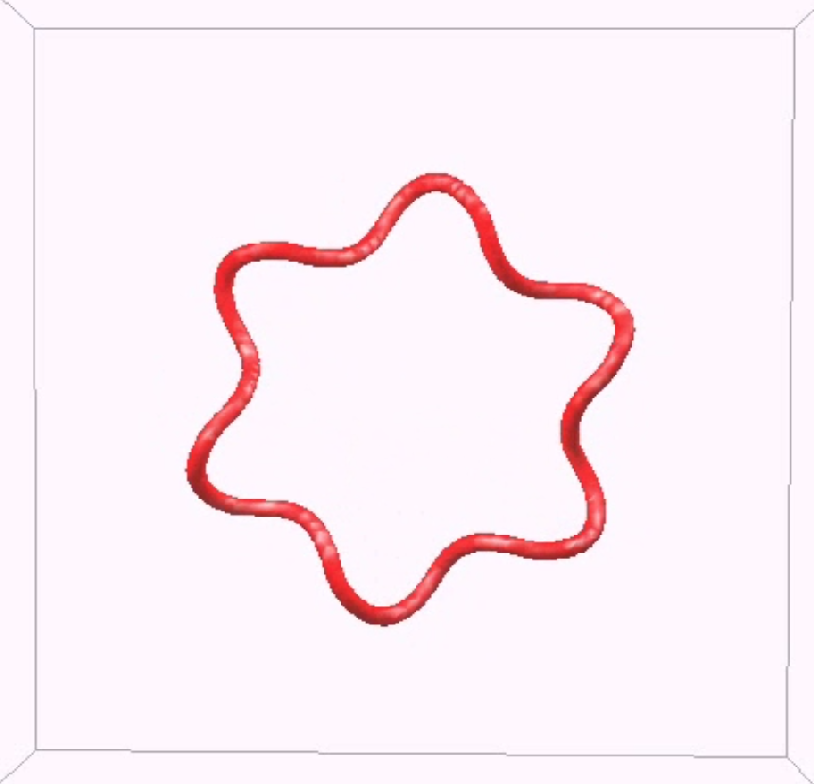}
\includegraphics[angle=0,width=0.32\textwidth]{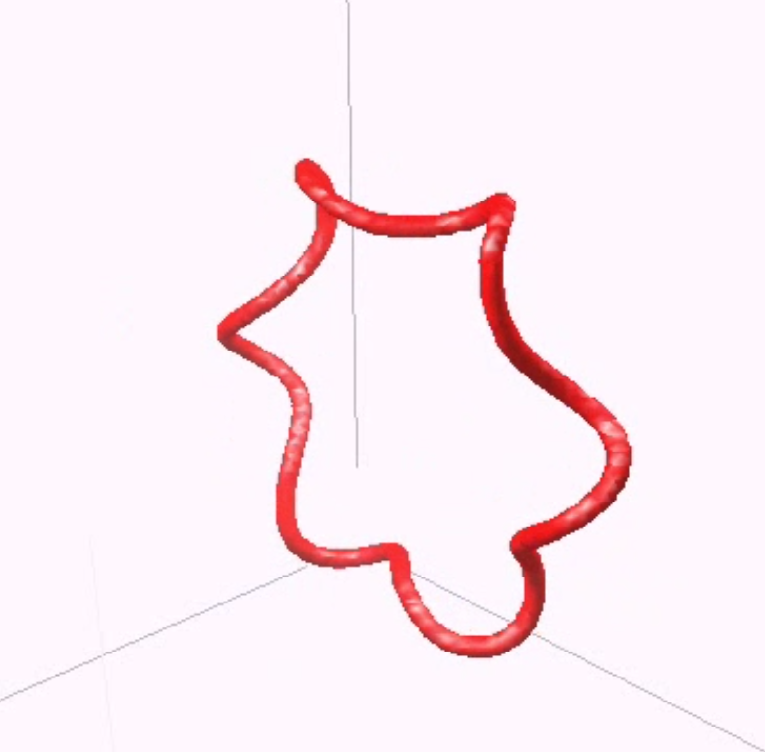}
\caption{Left: schematic Kelvin wave of wavelength $\lambda$
on a straight vortex line. The red line
is the perturbed vortex core, the dashed black line is the location of
the unperturbed straight vortex.
Middle and right:
a vortex ring, perturbed by a helical Kelvin wave having
$m=6$ azimuthal symmetry, is shown from two different directions.
Calculation performed using the GPE.}
\label{fig:Kelvin}
\end{figure}

Vortex lines sustain rotating helical perturbations of their shape
called {\it Kelvin waves}. Fig.~\ref{fig:Kelvin} (left) shows a Kelvin wave
of wavelength $\lambda$ on a straight vortex; if the amplitude of the 
wave is small, the angular velocity of rotation is proportional 
to $\lambda^{-2}$ (i.e. shorter waves rotate faster). 
Fig.~\ref{fig:Kelvin} (middle and right)
shows a vortex ring with a Kelvin wave of
azimuthal symmetry $m=5$. If the amplitude of
Kelvin waves on a vortex ring is large, the translational velocity
of the ring is reduced compared to a circular ring \cite{Hanninen}.
In general, the larger the temperature the smoother the vortex lines
become, as friction damps out the Kelvin waves and cusps caused by
reconnection events.

\subsection{Sound emission}
\label{sub:sound}

\begin{figure}[!ht]
\centering
\includegraphics[angle=0,width=0.25\textwidth]{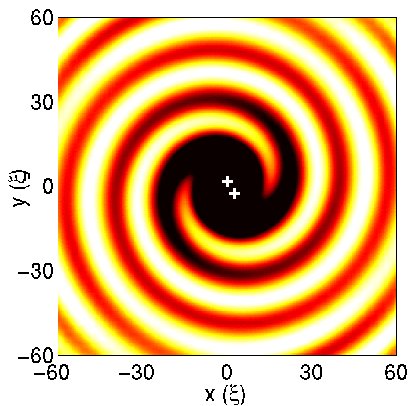}
\includegraphics[angle=0,width=0.27\textwidth]{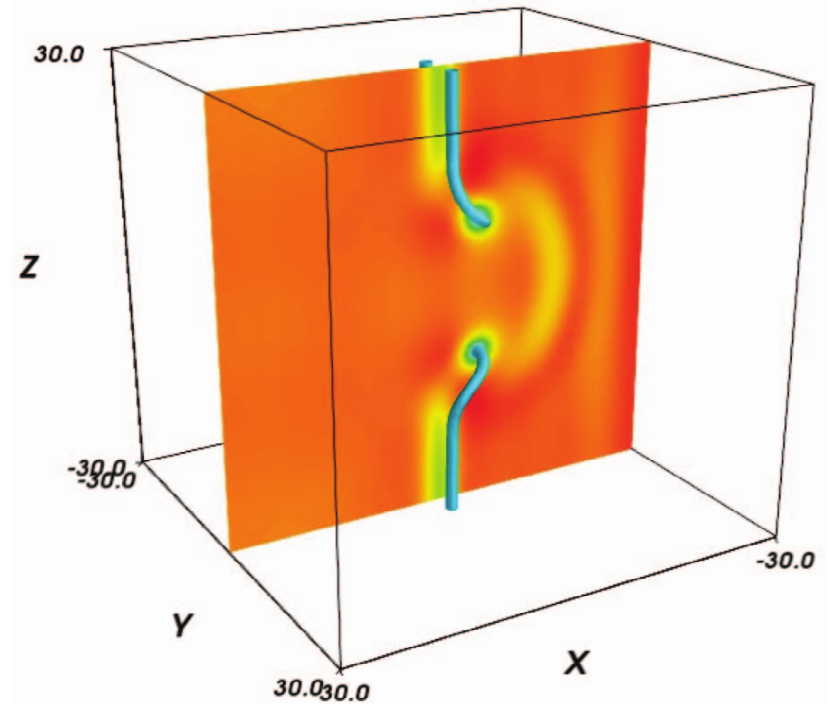}
\caption{
Left: Density wave pattern generated by a rotating vortex-vortex pair at the
centre of the figure (marked by the two white crosses). 
Bright/dark colours represent
density which is larger/smaller than the unperturbed density
(yellow). Calculation performed using the GPE \cite{Parker}.
Right: Final stage of the antiparallel reconnection shown 
in Fig.~\ref{fig:Zuccher}: note the strong sound pulse on
the plane between the two vortex lines, created at the
reconnection event. Calculation performed 
using the GPE \cite{Zuccher}.}
\label{fig:sound}
\end{figure}

Kelvin waves radiate sound waves (density oscillations)
as they rotate \cite{Vinen2001}.
In general, when a vortex line changes its direction
of motion (i.e. it accelerates), it loses energy by radiating 
sound waves \cite{Parker}, as is shown in Fig.~\ref{fig:sound}
(left) for a rotating vortex-vortex pair. This mechanism of dissipation of
kinetic energy is important in the case of short, rapidly rotating waves, 
and at low temperatures, for which friction with the normal fluid
is negligible: it explains the observed decay of turbulence at
temperatures as low as few mK \cite{Svistunov}. 
Similarly, vortex reconnections (see Section~\ref{sub:recon})
create a sound wave \cite{Leadbeater,Zuccher}, 
as is shown in Fig.~\ref{fig:sound} (right),
contributing to the energy dissipation.

\subsection{Reconnections}
\label{sub:recon}

\begin{figure}[!h]
\begin{center}
\includegraphics[angle=0,width=0.40\textwidth]{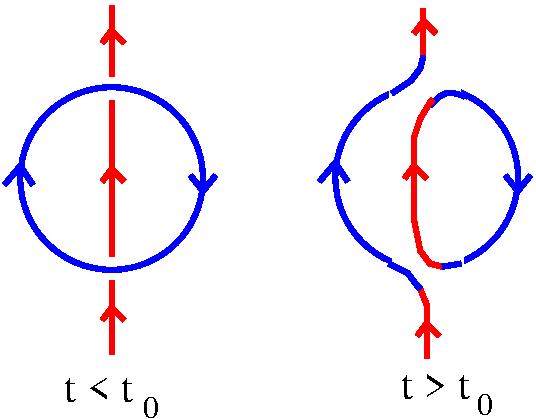}
\caption{Schematic collision between
a vortex ring (travelling into the page)
and a stationary straight vortex line, as is also shown in Fig.~\ref{fig:Youd}.
The arrows denote the direction of the vorticity. The figure shows
that, after the reconnection at time $t=t_0$, part of
the vortex ring has becomes part of the vortex line and vice versa.}
\label{fig:Youd-schematic}
\end{center}
\end{figure}

\begin{figure}[!ht]
\centering
\includegraphics[angle=0,width=0.60\textwidth]{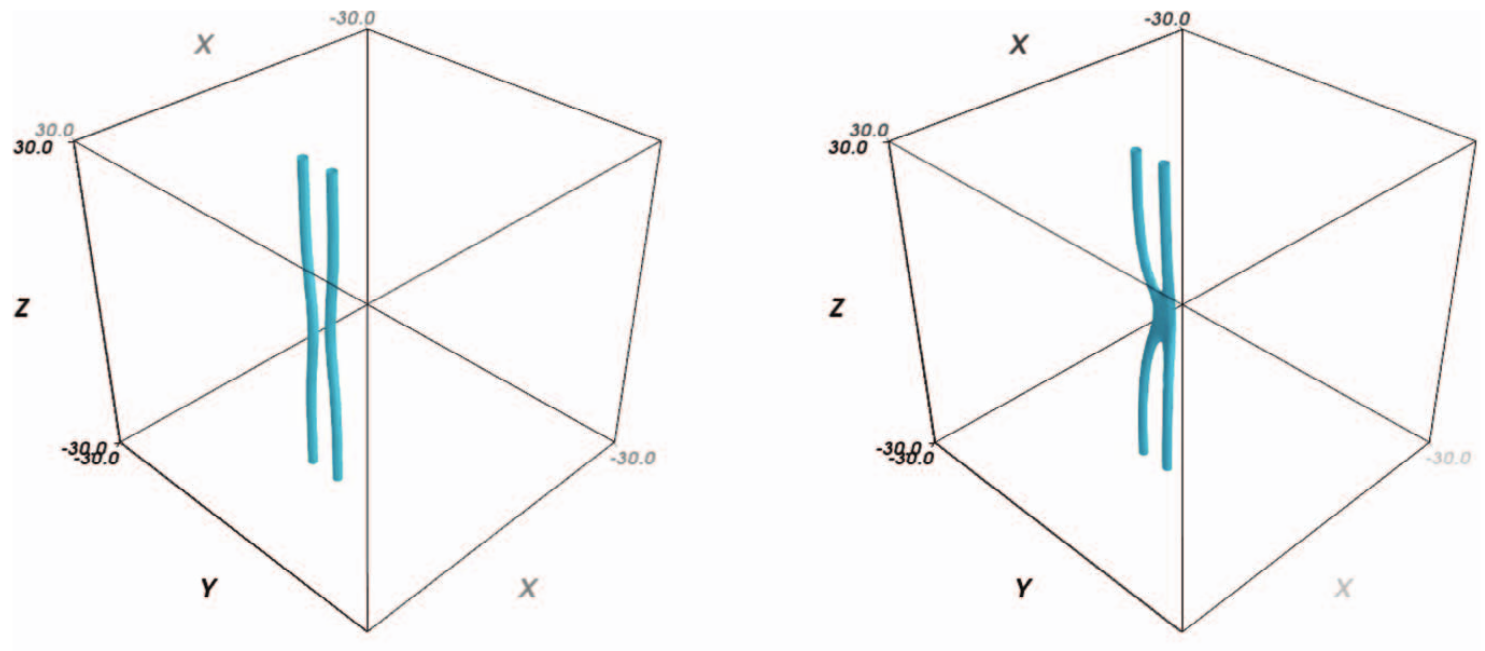}\\
\includegraphics[angle=0,width=0.60\textwidth]{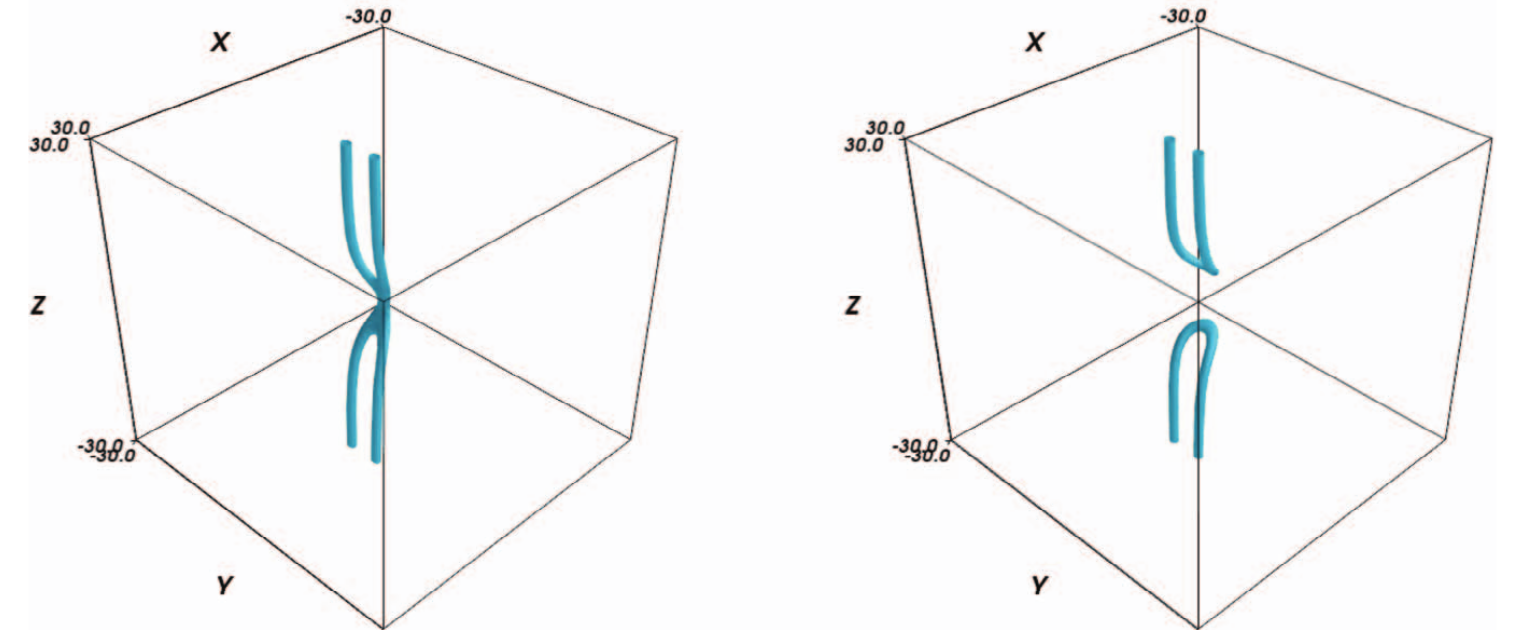}
\caption{Reconnections of two antiparallel vortex lines computed using
the GPE \cite{Zuccher}. 
Initially ($t<t_0$, top left) the vortices are slightly curved in order
to induce a motion which will result in a reconnection at $t=t_0$ (top right)
when the vortex cores merge; the two bottom figures show the reconnected
vortex lines which move away from each other.}
\label{fig:Zuccher}
\end{figure}

\begin{figure}[!h]
\centering
\includegraphics[angle=0,width=1.10\textwidth]{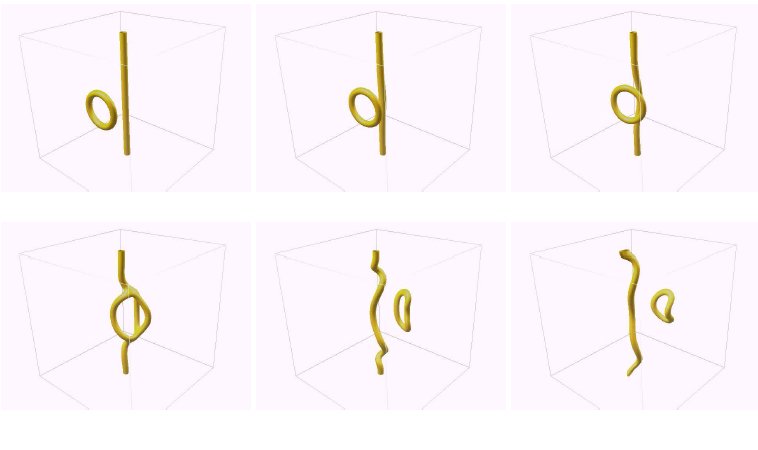}
\caption{Collision of a small vortex ring travelling from left to
right towards a straight vortex line computed
using the condensate model. The same collision is
represented schematically in Fig.~\ref{fig:Youd-schematic}.
Density isosurfaces 
visualize the vortex cores. Note that after the collision
the vortex ring has developed an $m=2$ Kelvin wave; the vortex line
has Kelvin waves too. Calculation performed using the GPE.}.
\label{fig:Youd}
\end{figure}

When vortex lines collide, they
reconnect \cite{reconnections}, as is shown schematically in 
Fig.~\ref{fig:Youd-schematic}.  The reconnection of two 
antiparallel vortex lines computed using the GPE
is shown in Fig.~\ref{fig:Zuccher}.  Individual
vortex reconnections have been experimentally observed both in helium~II and in
atomic condensates. Reconnections change the topology of the
vortex configurations; they are physically significant  because 
part of the kinetic energy of the colliding vortices is turned into
a sound pulse \cite{Leadbeater}. 
Some of the initial energy is also used to create 
Kelvin waves, as is shown in Fig.~\ref{fig:Youd}: here
a vortex ring travels from left to right towards a straight vortex line,
and collides with it, causing two reconnection events. 
It is apparent from the figure that, after the collision,
the vortex ring has acquired an $m=2$ Kelvin wave; the vortex line
exhibits Kelvin waves too.
Fig.~\ref{fig:Youd-schematic} represents the collision
schematically. If two vortex rings of the same size collide head-on,
we have an {\it annihilation}, and the entire kinetic energy of the
two vortices is turned into sound \cite{Leadbeater}.

\subsection{The Vortex Filament Model}
\label{sub:vfm}

There is not a single model which describes the phenomenology
of turbulent vortices in superfluid helium. 
The range of physics which is accessible in helium~II
in the same experiment is extreme. For example, at high temperatures
near $T_{\lambda}$ the mean 
free path of thermal excitations
is smaller than the vortex core, but at low temperatures it is
larger than the size of the typical 
apparatus \cite{Morishita}. In ordinary fluids,
these two limits are described by different models, the Navier-Stokes
equation and the Boltzmann equation respectively.

There are two main models which have been used to study
quantum turbulence: the GPE and the Vortex Filament Model (VFM). 
The advantage of the GPE is that it accounts for sound waves, vortex
nucleation and vortex reconnections. The disadvantage is that it does
not include the normal fluid and friction, which
are well accounted by the VFM. However the VFM does not include waves
and vortex reconnections (which must be performed algorithmically);
moreover, the VFM does not include nucleation, so the initial condition 
typically consists of some seeding vortices which  
can lengthen and propagate during the evolution.

\begin{figure}[!ht]
\scalebox{0.31}{\includegraphics{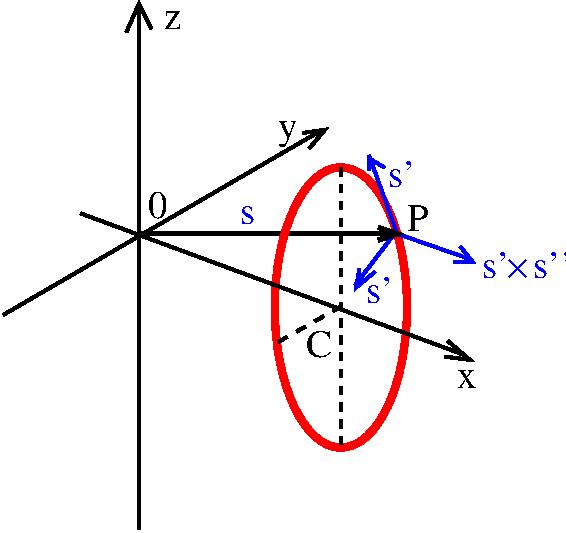}}~~~~~~~~~~~~~
\scalebox{0.28}{\includegraphics{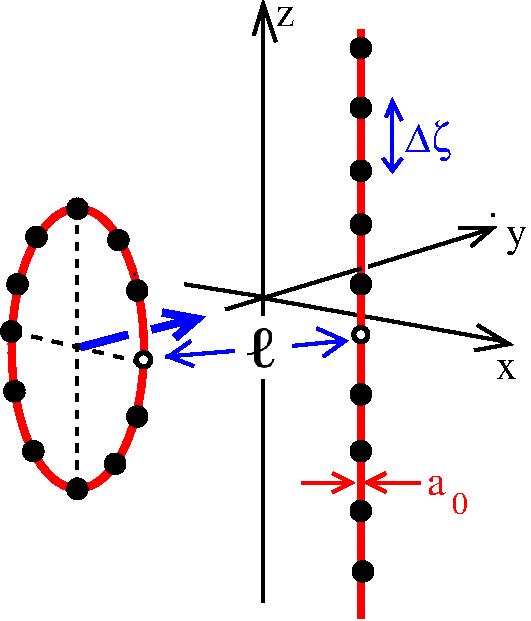}}
\caption{The Vortex Filament Method.
Left: Vortex ring (red line)
travelling in the $x$-direction. The plane of the ring
is parallel to the $yz$-plane and its centre is the point $C$.  At
a point $\bolds=\vec{OP}$ along the ring, the
three vectors $\bolds'$, $\bolds''$ and 
$\bolds' \times \bolds''$ are displayed as blue arrows.
Right: a vortex ring (whose plane is parallel to the $xz$-plane)
travels in the $y$-direction approaching a straight
vortex line parallel to the $z$-axis. The discretization points along the
ring and the line are marked by black circles. The figure also
shows the discretization
distance, $\Delta \zeta$, between the discretization points, the
vortex core radius, $a_0$, and
the minimum distance between the ring and the line, $\ell$ (the
discretization points used to define $\ell$ are marked as white circles).}
\label{fig:vfm}
\end{figure}

The VFM is built on the fact that in helium~II 
the vortex core radius, $a_0\approx 10^{-10}~\rm m$, 
is many orders of magnitude smaller than any length scale of the flow.
The VFM describes vortex lines as space curves of infinitesimal thickness
which are represented parametrically as $\bolds=\bolds(\zeta,t)$
where $\zeta$ is the arc length. At each point $\bolds$ along a vortex,
three orthogonal vectors are defined in the tangent, normal and binormal
directions respectively:
$\bolds'=d\bolds/d\zeta$, 
$\bolds''=d^2 \bolds/d\zeta^2$ and
$\bolds' \times \bolds''$ (where a prime denotes the derivative with
respect to arc length), as is shown for a vortex ring in Fig.~\ref{fig:vfm}.

Consider a vortex tangle $\cal T$ consisting of
many such space curves. The space curves are discretized by a large
number of points separated by the distance $\Delta \zeta$, representing
the numerical resolution of the calculation. The
discretization distance must be much larger than the expected
average separation between the vortex lines, $\ell \approx 10^{-3}$ to
$10^{-5}~\rm m$ in typical experiments. Clearly it would be computationally
impossible to reduce $\Delta \zeta$ to the dimensions of $a_0$,
so all phenomena taking place at length scales between $\Delta \zeta$
and $a_0$ must be ignored. 

It can be shown that, in the absence of any superfluid potential flow which
is imposed externally, the point $\bolds$ along a vortex line moves with
velocity \cite{Schwarz}

\begin{equation}
\frac{d\bolds}{dt}=\boldv_s+\alpha \bolds' \times (\boldv_n-\boldv_s)
-\alpha' \bolds' \times (\bolds' \times (\boldv_n-\boldv_s)),
\end{equation}

\noindent
({\it Schwarz's equation})
where $\boldv_n$ is any normal fluid velocity which is imposed externally and
$\alpha$, $\alpha'$ are temperature-dependent friction coefficients.
The superfluid velocity $\boldv_s$ at the point $\bolds$ is decomposed as

\begin{equation}
\boldv_s(\bolds,t)=\boldv_{loc}(\bolds,t)+\boldv_{non}(\bolds,t),
\label{eq:locnon}
\end{equation}

\noindent
where $\boldv_{loc}(\bolds,t)$ arises from the 
local curvature at the point $\bolds$

\begin{equation}
\boldv_{loc}(\bolds,t)
= \frac{\kappa}{4 \pi} \ln{\left(\frac{R}{a_0}\right)}
\bolds' \times \bolds'',
\label{eq:loc}
\end{equation}

\noindent
and $\boldv_{non}(\bolds,t)$ accounts for
non-local velocity contributions due to (i) other parts of the same vortex line
and (ii) other vortex lines in $\cal T$

\begin{equation}
\boldv_{non}(\bolds,t)=
-\frac{\kappa}{4 \pi} \oint_{\cal T'}
\frac{(\bolds-\bolds_0(\zeta))}{\vert \bolds-\bolds_0(\zeta) \vert^3}
\times \bolds'_0(\zeta) d\zeta,
\label{eq:far}
\end{equation}

\noindent
({\it Biot-Savart law}) where $\cal T'$ is the vortex tangle $\cal T$ 
excluding the close neighborhood of the point $\bolds$, thus preventing
the integral from diverging.
Notice that the local contribution, $\boldv_{loc}(\bolds,t)$,
points in the binormal direction and is inversally proportional
to the radius of curvature of the vortex line at $\bolds$, defined as
$R = 1 / \vert \bolds'' \vert$. Vortex reconnections are implemented
algorithmically when two vortex strands become closer than the
minimum distance $\Delta \zeta$.

The friction coefficients $\alpha$ and $\alpha'$  
tend to zero for $T \to 0$. In this limit of zero
temperature, Schwarz's equation reduces to

\begin{equation}
\frac{d\bolds}{dt}=\boldv_s=\boldv_{loc}+\boldv_{non},
\end{equation}

\noindent
which restates the classical Helmholtz's theorem that
vortex lines move with the (super)flow.

\section{TOPOLOGY OF QUANTUM TURBULENCE}
\label{sec:topology}

Individual {\it vortex knots} and their stability have been
studied for some time \cite{Kida,Ricca-Samuels-Barenghi,Proment,Oberti}.
Images such as Fig.~\ref{fig:tangles} raise the question:
are vortex lines in quantum turbulence knotted? If they are,
what is the amount of this "knottiness"? 
If the turbulence is sustained in a statistical steady-state,
does the knottiness fluctuate about a well-defined average value?
Is this average value related to the
intensity of the turbulence, e.g. the Reynolds number? 
Vortex knots can be created (and destroyed) only by reconnections,
events which are associated with dissipation of kinetic energy.
Therefore, a knotted vortex
tangle which decays must release a certain amount of energy in the
form of sound waves simply by reducing its topological complexity.
Another question is whether this decay takes place along preferred
topological pathways, as has been suggested 
\cite{Kleckner-untie,LiuRicca2016,LiuRiccaLi}.
These questions have acquired more interest since Irvine and
collaborators showed how to create vortex knots in the laboratory
in a controlled, reproducible way \cite{Kleckner-knot}. 

\subsection{Steady turbulence in open domain}
\label{sub:creating}

The first step towards answering these questions is to quantify 
the knottiness of vortex tangles in ways that can be
computationally implemented, ideally even in extremely
dense vortex tangles such as
Fig.~\ref{fig:Baggaley-EPL}.  
All calculations presented here refer to vortex tangles which
evolve in an infinite computational domain. This is because
periodic boundary
conditions (used to study homogeneous isotropic turbulence) would 
complicate the definitions of the measures of complexity,
which would not be helpful at this early stage of investigation.

Clearly it would be preferable to study regimes in which the turbulence 
is in a statistical steady state, so that the average properties of the 
turbulence can be related to how the turbulence is driven. 
At nonzero temperature, a statistical steady state 
requires a continuous input of energy by the driving normal fluid, 
which in an experiment would be imposed externally. 
The results which are presented here refer to two
choices for the driving normal fluid velocity $\boldv_n$.
The first choice consists of random waves \cite{Mae} (exponentially 
localized in a spherical region at the
centre of the computational domain). The second choice \cite{Cooper} is a
Dudley-James flow. This flow (used in numerical models of planetary dynamos)
is both solenoidal and localized in a sphere of radius $D$; it has the form

\begin{figure}[!ht]
\centering
\includegraphics[angle=0,width=0.50\textwidth]{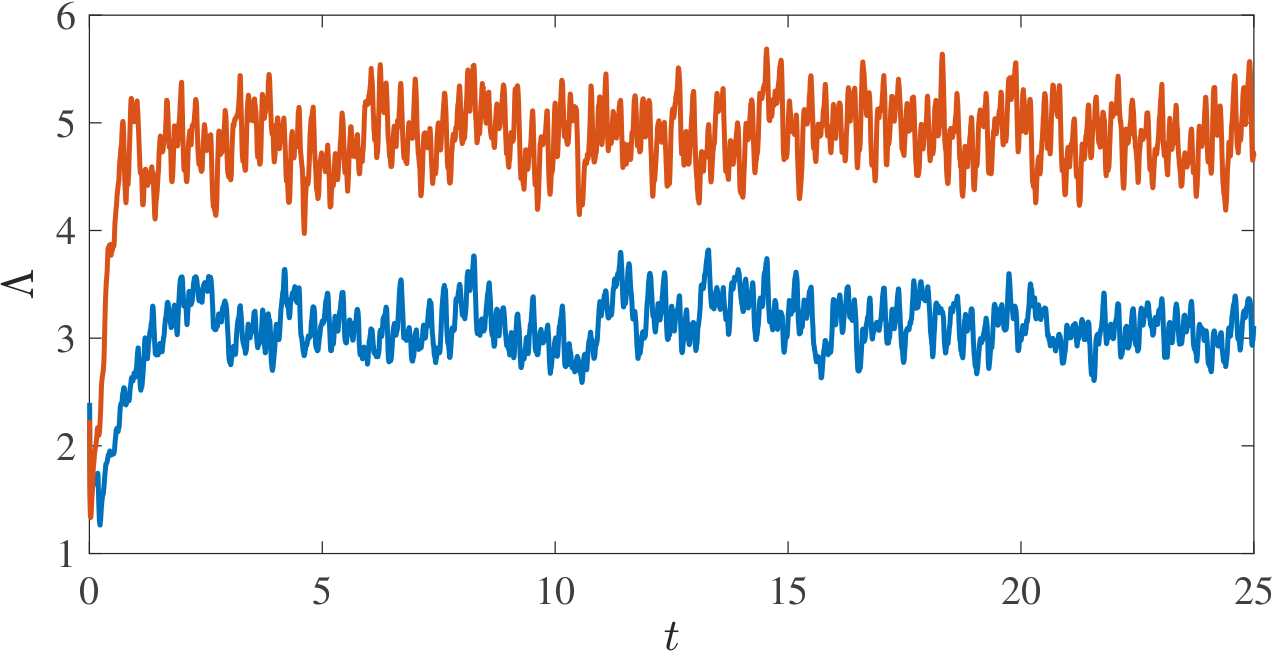}
\caption{Total vortex length, $\Lambda$, vs time, $t$, for two different
driving normal fluid velocities $\boldv_n$. The calculation starts with
some seeding vortex lines. After an initial transient,
the total vortex length $\Lambda$
saturates to an average value which does not depend on the
initial condition (the larger $\boldv_n$, the larger $\Lambda$). 
From Ref.~\cite{Cooper}.}
\label{fig:Rob-fig2}
\end{figure}

\begin{figure}[!ht]
\centering
\includegraphics[angle=0,width=0.60\textwidth]{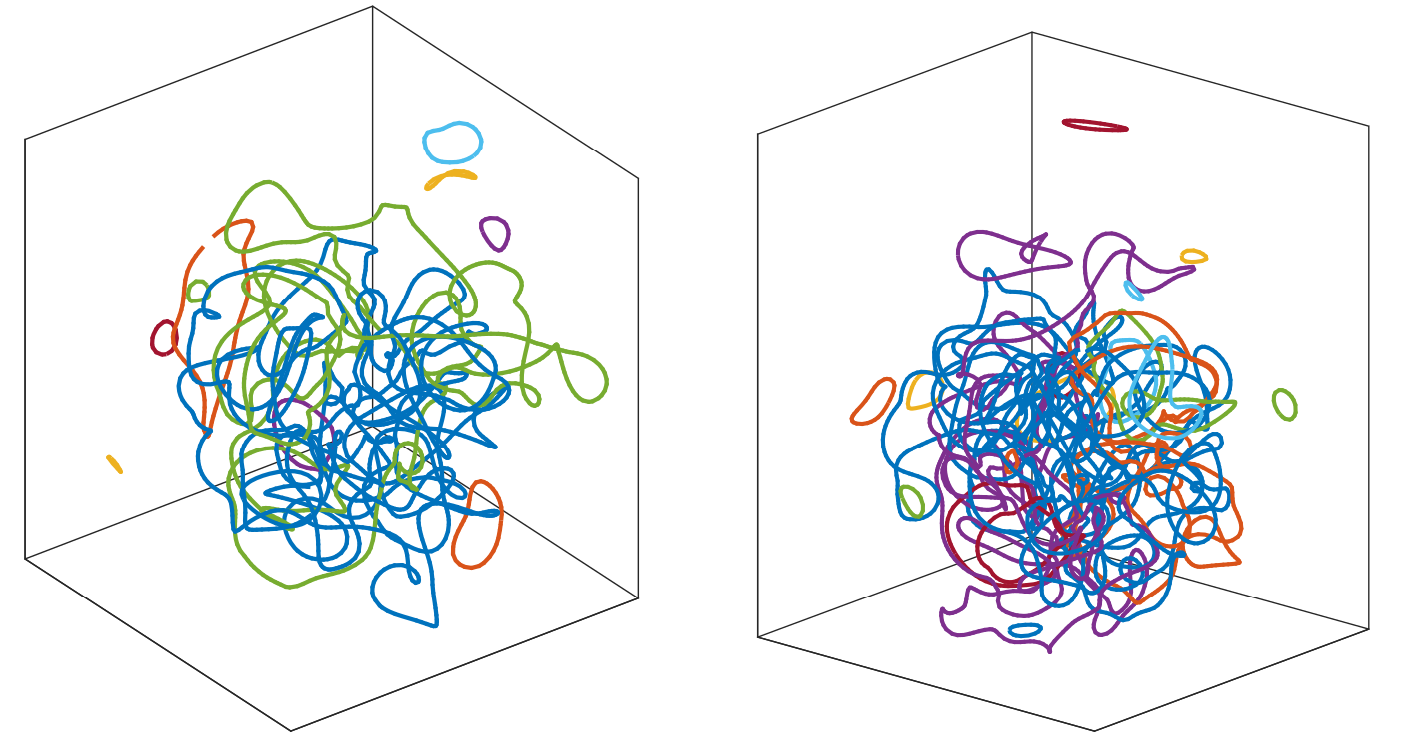}
\caption{Snapshots of vortex tangles in the saturated statistic steady-state
regime corresponding to the two driving normal fluid velocities
of Fig.~\ref{fig:Rob-fig2}. Each vortex loop is colour-coded differently. 
From Ref.~\cite{Cooper}.}
\label{fig:Rob-fig3}
\end{figure}

\begin{equation}
\boldv_n(r,\theta,\phi)=\sum_{l,m} (\boldt^m_l+\bolds^m_l),
\end{equation}

\noindent
where $r,\theta,\phi$ are spherical coordinates,

\begin{equation}
\boldt^m_l=\nabla \times \rhat \, t^m_l Y^m_l(\theta,\phi),
\quad
\bolds^m_l=\nabla \times \nabla \times \rhat \, s^m_l Y^m_l(\theta,\phi),
\quad
-l \le m \le l,
\end{equation}

\noindent
and $Y^m_l$ are spherical harmonics. Results presented here refer to
$m=0$, $l=2$, $t^0_2=s^0_2=r^2 \sin{(\pi r/D)}$ \cite{Cooper}.

The results obtained with these choices of $\boldv_n$ 
are qualitatively similar.
By forcing the turbulence in the central region only,
a balance is achieved in this region between forcing and friction. 
The occasional vortex loop which drifts out of
the central region and moves into quiescent normal fluid decays.
A statistically steady-state of
turbulence consisting of loops is thus created: hereafter
we use the symbol ${\cal T}=\bigcup_{i=1}^N {\cal L}_i$ to mean
that the vortex tangle $\cal T$ consists of a collection of $N$ closed
vortex loops ${\cal L}_j$ ($j=1, \cdots, N$); the number $N$
fluctuates around a mean value because of continual vortex
reconnections.

Fig.~\ref{fig:Rob-fig2} confirms that
after an initial transient, the total length of vortex
lines, $\Lambda$, achieves well-defined average value in the
statistical steady state regime.
Fig.~\ref{fig:Rob-fig3} shows two snapshots of vortex tangles 
corresponding to the small and large values of $\boldv_n$ of
Fig.~\ref{fig:Rob-fig2}. The energy of the vortex tangle, defined as

\begin{equation}
E=\frac{1}{2} \rho_s \int \boldv(\boldr) \cdot \boldv(\boldr) \, d^3\boldr
=\rho_s \kappa \oint_{\cal T} \boldv_s \cdot \boldr \times \bolds' d\zeta,
\end{equation}
\noindent
has the same time behaviour of $\Lambda$
shown in Fig.~\ref{fig:Rob-fig2}, fluctuating
about a well-defined average value after the initial transient.

\subsection{Quantifying the complexity of the turbulence}
\label{sub:quantifying}

\begin{figure}[!h]
\centering
\includegraphics[angle=0,width=0.80\textwidth]{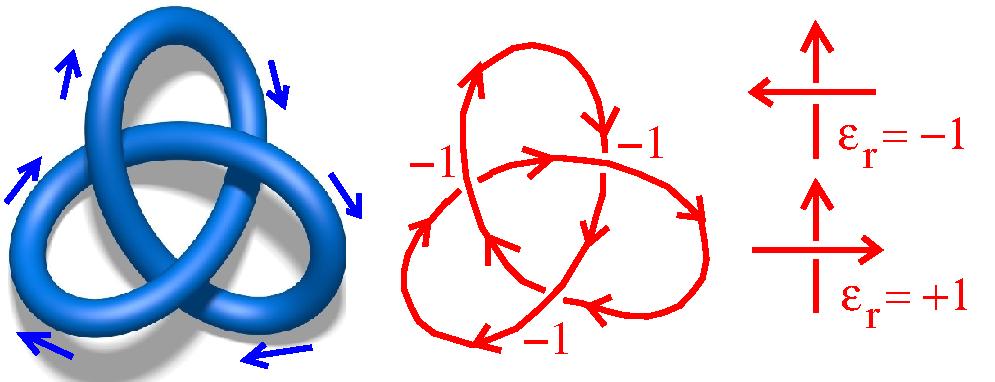}
\caption{Left: a  vortex in the shape of a trefoil knot; the
tube represents the region of depleted density around the axis of the
vortex. The axis of the tube is a three-dimensional space curve whose
orientation is provided by the direction
of the vorticity, that is, by the clockwise
or anticlockwise direction of motion of the fluid around the vortex axis.
Middle: The three-dimensional space curve is projected onto an
arbitrary plane, making visible three apparent intersections; the
two-dimensional curve in the figure
is interrupted, to retain the information about which
strand is above and which strand is below in the actual three-dimensional
configuration on the left.
Right: we follow the two-dimensional curve starting from an arbitrary
point. At each apparent intersection, $r$, if the other vortex strand is above,
we define a crossing number $\epsilon_r=\pm 1$ depending on the other
strand pointing right or left (if the other strand is below, we ignore 
this intersection and move to the next). In this
example we find $\epsilon_r=-1$ for all $r=1,2,3$, hence obtain the writhing
number ${\rm Wr}=-3$.}
\label{fig:crossing}
\end{figure}

A possible strategy \cite{Barenghi-Ricca-Samuels} is to operate with 
{\it crossing numbers}, as is illustrated in Fig.~\ref{fig:crossing}: 
consider a trefoil vortex loop (left), a three-dimensional, closed 
space curve oriented by the direction of the vorticity (indicated by arrows). 
We project
the curve onto an arbitrary two-dimensional plane (middle) and assign
the  crossing number $\epsilon=\pm 1$ to each apparent
intersection in a standard way (explained in the figure caption). 
Interesting geometrical and topological properties can be
assembled from these crossing numbers.
For a single loop ${\cal L}_i$, the {\it writhing
number}, ${\rm Wr}({\cal L}_i)$, provides a geometrical measure 
of the coiling. The definition is

\begin{equation}
{\rm Wr}({\cal L}_i)=\Big \langle 
\sum_{r \in {\cal L}_i \cap {\cal L}_i} \epsilon_r 
\Big \rangle,
\label{eq:writhing}
\end{equation}

\noindent
where $\langle \cdots \rangle$ denotes averaging over all directions
of projection.
A measure of the total coiling of the entire vortex tangle
is therefore

\begin{equation}
{\rm Wr}({\cal T})=\Big \langle 
\sum_{r \in {\cal T}} \epsilon_r
\Big \rangle.
\end{equation}

\noindent
In practice, numerical experiments \cite{Barenghi-Ricca-Samuels} have
shown that, provided the vortex tangle is sufficiently dense and
isotropic, averaging over all projections is well approximated by 
the simpler average over the three Cartesian directions. 

Another measure of the complexity of a vortex tangle is the
{\it average crossing number}, representing
the average number of apparent unsigned crossings. 
For two loops ${\cal L}_i$ and
${\cal L}_j$, this quantity is defined as

\begin{equation}
C({\cal L}_i,{\cal L}_j)=\big \langle
\sum_{r \in {\cal L}_j \cap {\cal L}_j} \vert \epsilon_r \vert
\big \rangle.
\label{eq:average-crossing-number}
\end{equation}

\noindent
The generalization to the entire vortex tangle is 

\begin{equation}
C({\cal T})=\sum_{{\cal L}_i, {\cal L}_j \in {\cal T}} 
C({\cal L}_i,{\cal L}_j).
\end{equation}

\begin{figure}[!ht]
\centering
\includegraphics[angle=0,width=0.80\textwidth]{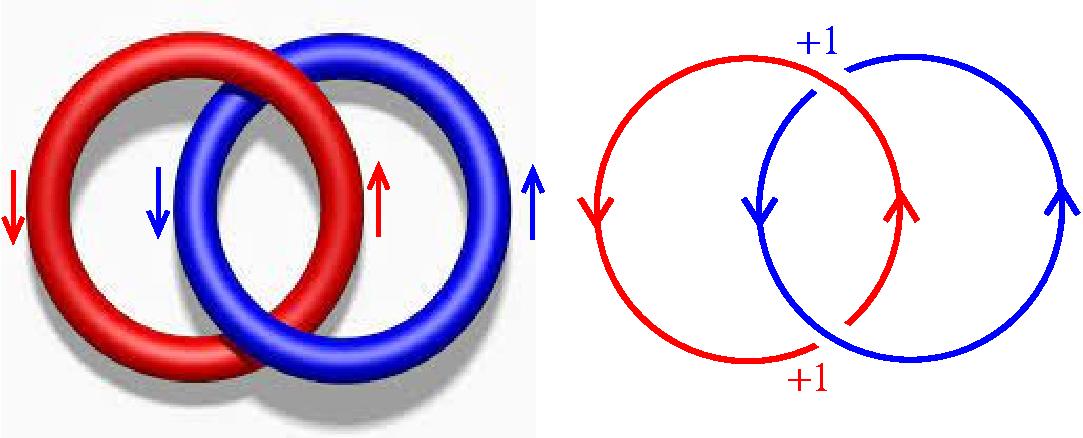}
\caption{Left: Hopf link.
Right: by applying Eq.~\ref{eq:link} we find that
the linking number of this vortex configuration is 
${\rm Lk}({\cal L}_1,{\cal L}_2)=(1/2)(1+1)=1$.}
\label{fig:link}
\end{figure}

The linking number between two loops ${\cal L}_i$ and
${\cal L}_j$ is one of the most important topological invariants 
(it does not change if the loops are deformed). It is defined 
in terms of crossing numbers by

\begin{equation}
{\rm Lk}({\cal L}_i,{\cal L}_j)=
\frac{1}{2} \sum_{r \in {\cal L}_i \cap {\cal L}_j} 
\epsilon_r.
\label{eq:link}
\end{equation}

\noindent
An example of the calculation of the linking number is shown in
Fig.~\ref{fig:link}, which shows the simplest linked structure: 
the {\it Hopf link} between two vortex loops.
Notice that the linking number does not depend on the single projection 
over which it is calculated. 
Using Eq.~\ref{eq:link}, the total linking number of the entire vortex tangle 
is naturally defined as

\begin{equation}
{\rm Lk}({\cal T})=\sum_{ {\cal L}_i, {\cal L}_j \in {\cal T}, i \neq j} 
\vert \; {\rm Lk}({\cal L}_i,{\cal L}_j) \; \vert,
\label{eq:linkage}
\end{equation}

\noindent
(note that contributions from self-linking are excluded). This 
quantity provides a simple global measure of the topological complexity of 
a turbulent system which can be computed relatively easy, being the 
vorticity discrete. It is important to notice that
writhing number, average crossing number and linking number can also be
calculated by performing Gauss integrals; however, it was found by
numerical experiments \cite{Barenghi-Ricca-Samuels} that, particularly
for the dense tangles of vortices which interest us, the
expressions in terms of crossing numbers are numerically more robust.
The reason is that the calculation involves only integers, unlike the Gauss 
integrals which depend on the discretization along the space curves.

\begin{figure}[!ht]
\centering
\includegraphics[angle=0,width=0.60\textwidth]{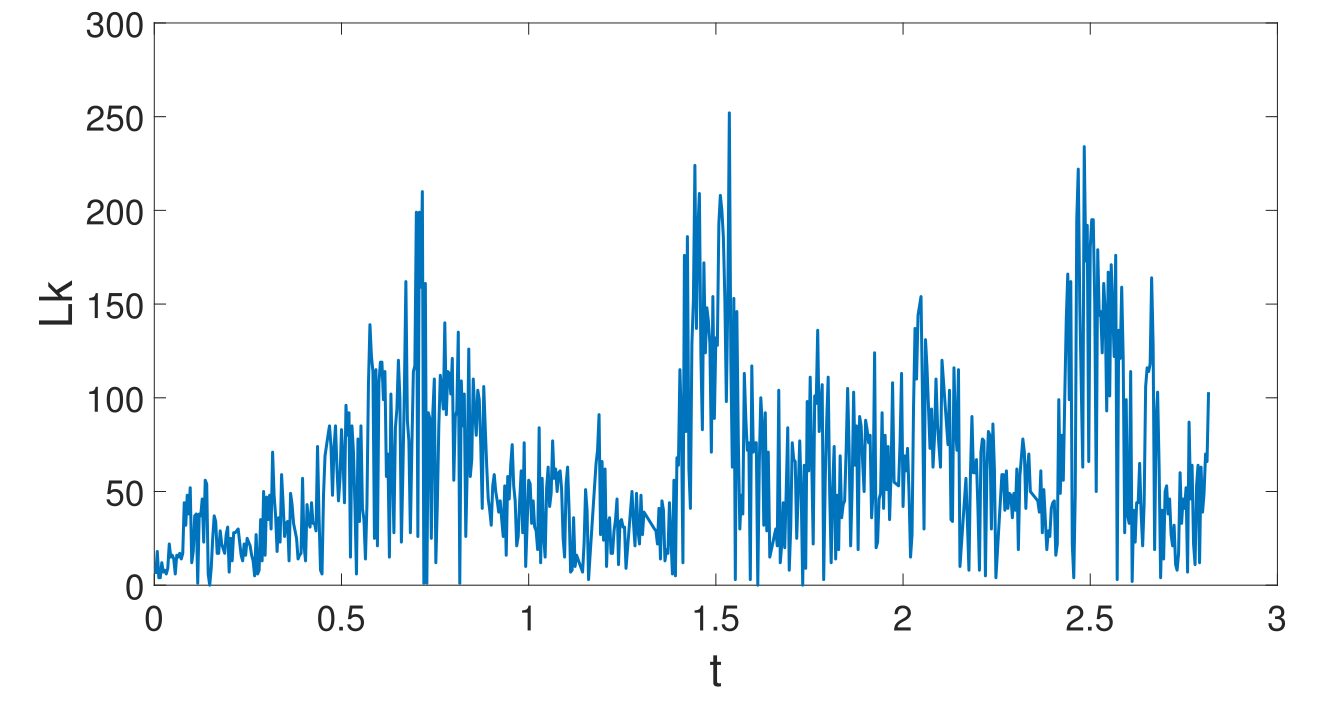}
\caption{Total linking number ${\rm Lk}$ vs time $t$
of a vortex tangle similar to Fig.~\ref{fig:Rob-fig2}.
From Ref.~\cite{Mae}.}
\label{fig:Mae-fig8}
\end{figure}

Numerical experiments confirm 
that the total linking number achieves a statistical steady state
\cite{Mae} after an initial transient, like the total length and the
energy, see Fig.~\ref{fig:Mae-fig8}.  The
fluctuations, caused by reconnections which continually split and merge vortex
loops, are relatively large for small vortex tangles, as expected.

\subsection{Helicity}
\label{sub:helicity}

Another quantity which can be interpreted as a measure of the
complexity of a turbulent flow is the {\it helicity} \cite{Moffatt1969},
a constant of motion under Euler evolution. In a classical
flow $\boldv$ (solution of the Euler or Navier-Stokes equation),
the definition of helicity is

\begin{equation}
\mathcal{H}=\int_V \bom(\boldr,t) \cdot \boldv(\boldr,t) \; d^3\!\boldr,
\label{eq:H1}
\end{equation}

\noindent
where $\bom(\boldr)=\nabla \times \boldv(\boldr)$ is the vorticity.
Unfortunately there is no consensus on what should be the
definition of helicity for a superfluid.  This problem is
mainly debated in the context of the GPE, for which vorticity is a delta
function centred on the vortex axis, where the azimuthal velocity
diverges. A definition of superfluid helicity based on Eq.~\ref{eq:H1}
thus requires \cite{diLeoni2017} a careful limit for $r \to 0$ (near the
vortex axis) and
opens the question whether $\cal{H}$ is conserved or not under GPE evolution.

A second approach
to superfluid helicity has been proposed based on the decomposition 
\cite{MoffattRicca1992} of $\mathcal{H}$ for thin vortex tubes into writhe, 
link and twist, in which case superfluid helicity is zero at all times
\cite{Hanninen2016,ZuccherRicca2018,Salman2018}.
However a subtle aspect arises:
the twist has an intrinsic component which requires 
the construction of a second line along the vortex axis in order to define
a ribbon; this is not a problem for a classical thin-cored vortex (however
small the core is, it contains infinite vortex lines), but runs into the
difficulty that, in the GPE model of the vortex core, there is only {\it one}
vortex line. 

A third proposal starts from the observation that the GPE only provides
a simplified model of the vortex core in helium~II.
A better model based on N-body quantum mechanics,
as explained in Section~\ref{sub:Reatto}, predicts that in the core region
the density does not vanish and the vorticity is uniform, thus
the classical definition of helicity, Eq.~\ref{eq:H1}, can be directly applied.
At the mesoscale level of description provided by the VFM (which neglects 
physics at scales between $a_0$ and $\Delta \zeta$), using 
Eq.~\ref{eq:locnon}, one finds \cite{Galantucci2021}

\begin{equation}
{\cal H}= \kappa \oint_{\cal T} \boldv_{non}(\bolds) \cdot \bolds' d\zeta
=\kappa \Lambda \langle \bolds' \cdot \boldv_{non} \rangle,
\end{equation}

\noindent
where $\langle \cdots \rangle$ denotes  the average over all vortex lines.
Thus $\cal H$ measures the nonlocal contribution to the lines'
velocities.  This result is interesting because, as mentioned in
Section~\ref{sec:dynamics}, two limiting cases of quantum turbulence
characterized by different energy spectra and decay laws have been
observed: Vinen turbulence and Kolmogorov turbulence.
In Vinen turbulence the vortex lines are randomly oriented, which
means that nonlocal contributions to the velocity of a line at a point
tend to cancel out and ${\cal H} \approx 0$. In Kolmogorov turbulence, 
the partially polarized bundles which contain most of the energy
and create the classical Kolmogorov spectrum give the necessary
nonlocal velocity contributions which keeps
$\cal H$ fluctuating around a nonzero value. This interpretation
of helicity is illustrated in Fig.~\ref{fig:tangles}:
the left and right tangles (corresponding to
large and small values of ${\cal H}$ respectively) are examples of Kolmogorov
and Vinen turbulence respectively.

\subsection{Knottiness}
\label{sub:Alexander}

\begin{figure}[!ht]
\centering
\includegraphics[angle=0,width=0.45\textwidth]{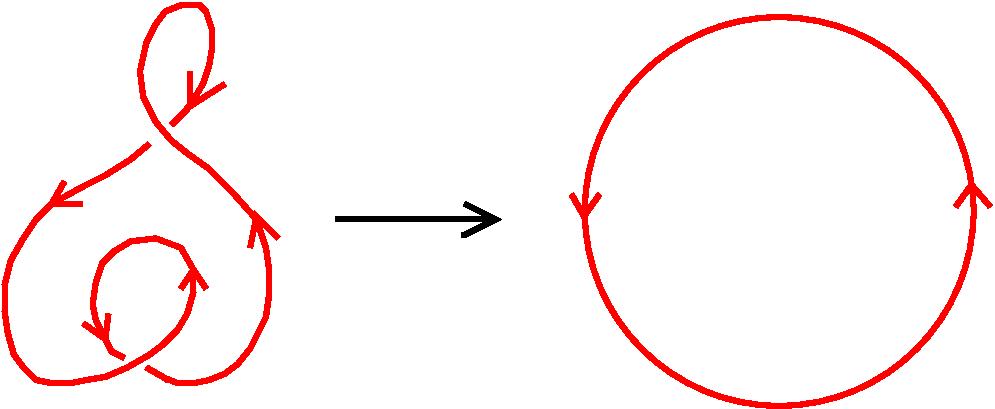}
\caption{This vortex loop is an unknot: it can be deformed into
a vortex ring without any reconnection. The arrows indicate the direction
of the vorticity.}
\label{fig:unknot}
\end{figure}

We have already seen that vortex tangles contain links.
Now we come to the question which motivates these investigations: 
are the vortex lines knotted in turbulent flows? 
To find the answer we need to take a 
three-dimensional snapshot of
the turbulence at a given time, consider
each vortex loop ${\cal L}_j \in {\cal T}$ at the time, and
numerically determine the knot type. The last step uses \cite{Cooper} 
the association between knot types and Alexander polynomials, written
in the form

\begin{equation}
\Delta(\tau)=c_0+c_1 \tau + c_2 \tau^2 + \cdots + c_{\nu} \tau^{\nu},
\end{equation}

\noindent
where $\nu$ is the degree of the polynomial,
and $c_0, \cdots c_{\nu}$ are integer
coefficients (in principle, other knot polynomials could be used). 
For example, a vortex ring has Alexander polynomial
$\Delta(\tau)=1$ which identifies the {\it unknot}: any closed vortex
loop which can be deformed into a ring without reconnections
(see Fig.~\ref{fig:unknot}) corresponds to $\Delta(\tau)=1$. The simplest
non-trivial knot is the
{\it trefoil knot}, which has Alexander polynomial
$\Delta(\tau)=1-\tau+\tau^2$ of degree $\nu=2$. In general, the
larger the degree $\nu$, the more complex the knot type. Any vortex
loop with degree $\nu>0$ is thus knotted (but the converse is not true,
counterexamples are known). The numerical algorithm to compute the
Alexander polynomials is based on finding crossings onto projections
\cite{Livingstone}
and is tested against all the knots of the Rolfsen table. Numerical
checks are applied by changing the discretization along the vortex
lines and by rotating the vortex loops, which changes the
number of crossings.
In this way, one finds the degree $\nu_j$ of all vortex loops
${\cal L}_j$ ($j=1, \cdots N$) for a given snapshot of the
turbulence, where the number of loops, $N$, changes with time,
as reconnections continually change, create
and destroy knots. Data are collected over time in the statistical
steady-state regime of turbulence and analyzed.

\begin{figure}[!ht]
\centering
\includegraphics[angle=0,width=0.60\textwidth]{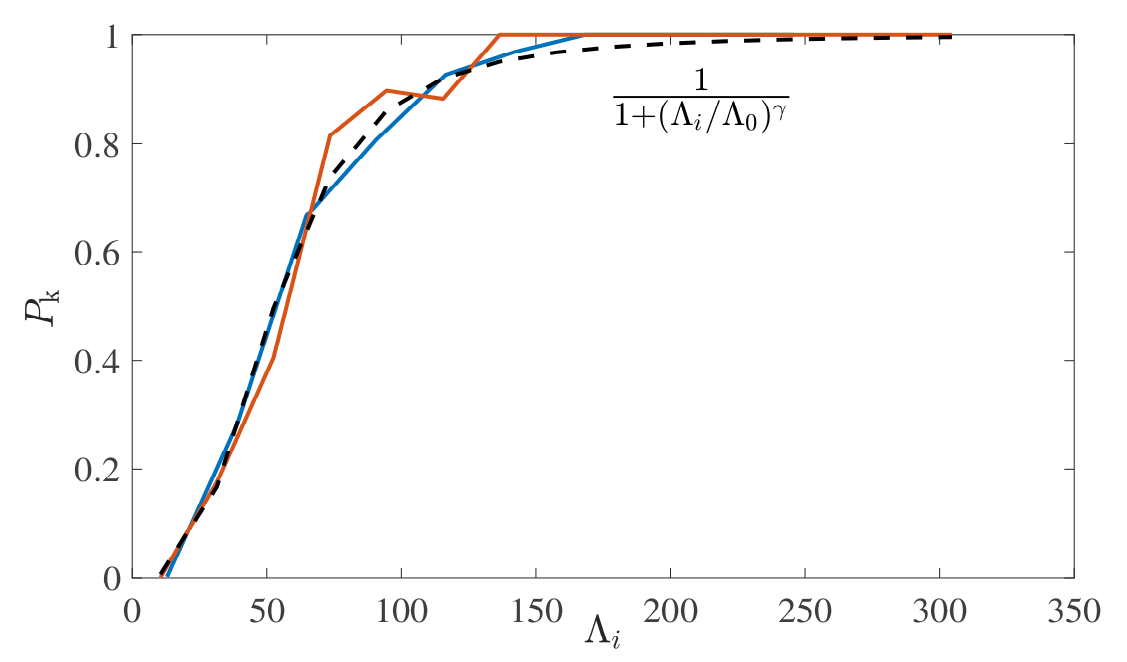}
\caption{Probability $P(\Lambda)$ that a vortex loop of length 
$\Lambda$ is knotted when
computed in the statistical steady state regime.
From. Ref.~\cite{Cooper}.}
\label{fig:Rob-fig5}
\end{figure}

\begin{figure}[!ht]
\centering
\includegraphics[angle=0,width=0.60\textwidth]{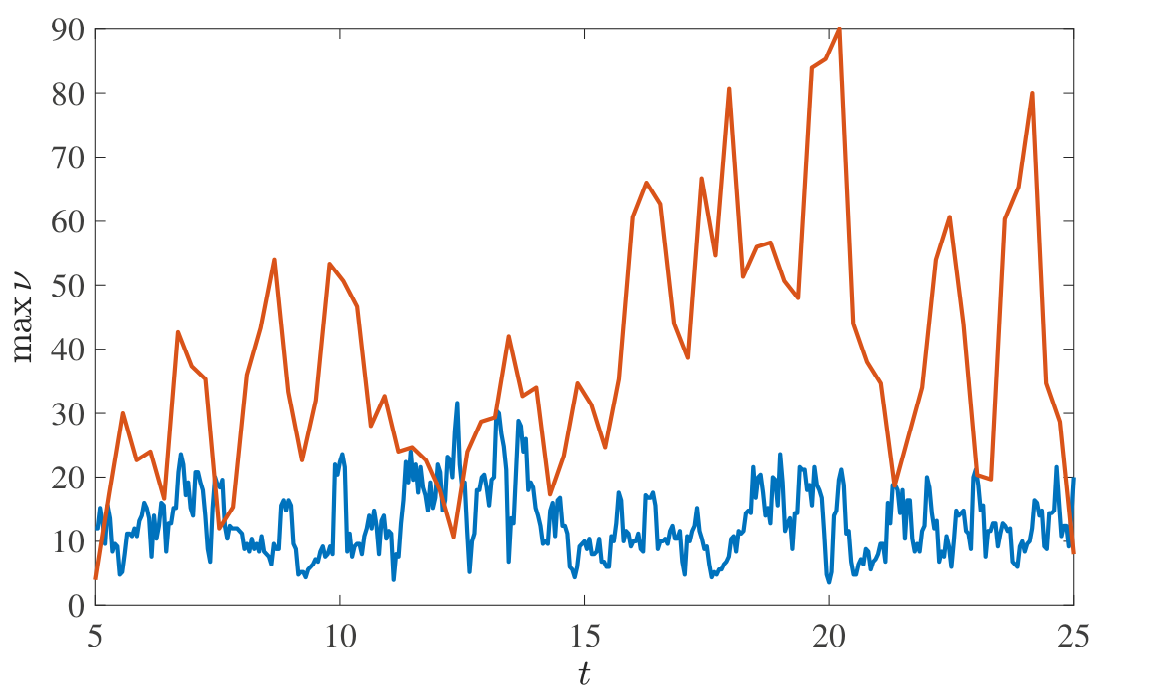}
\caption{The largest (red) and the second largest (blue) degree
of Alexander polynomial vs time $t$ in the statistical state state regime.
From Ref.~\cite{Cooper}.}
\label{fig:Rob-fig7}
\end{figure}

It is found \cite{Cooper} that the probability $P(\Lambda)$
that a vortex loop is knotted tends
to unity if the loop is sufficiently long, see Fig.~\ref{fig:Rob-fig5}.
This result is consistent with Monte Carlo simulations of random knotting
of DNA molecules in confined volumes \cite{Arsuaga}, and with experiments
in which strings of different lengths were tumbled inside a box
\cite{Raymer} (although in this case $P(\Lambda)$ does not tend to 
unity at large $\Lambda$, perhaps due to the stiffness of the strings).

What is remarkable is that, at all times, the vortex tangles contains
some vortex loops with very large degree of Alexander polynomial.
Fig.~\ref{fig:Rob-fig7} shows the two largest degrees as a function
of time in the statistical steady state regime: even for the small vortex
tangles investigated, degrees from $\nu=40$ to $90$ are not uncommon, 
and the second largest degree is typically around $\nu=10$ \cite{Cooper}. A
similar result is if found if the turbulence is driven by random waves
\cite{Mae}. Averaged
over a time window in the statistical steady state, the probability
that the largest degree of Alexander polynomial is $\nu$ appears 
to scale as $\nu^{-1.5}$.

\section{Conclusions}

At this point we can give a tentative answer to the question raised in
Section~\ref{sec:motivation}. We have seen that quantum turbulence is,
in some sense, the skeleton of classical turbulence: thin (microscopic)
filaments carrying one quantum of circulation each moving in an ideal fluid
background. We have seen that 
if we average their properties over a sufficiently large length
scale, the main (macroscopic) properties of classical turbulence emerge.
We have found that this turbulence is not only linked and knotted, but
it contains very high degree vortex knots. Monsters are lurking in the
vortex tangles.

The other main conclusion for the audience of these lectures is that, if the
temperature is reduced to almost absolute zero, matter does not necessarily
become a dull frozen solid, but can flow in extraordinary ways as a quantum
fluid: effects of quantum mechanics appear at the macroscopic level.

The computational tools which I have presented to quantify the disorder
of vortex lines have helped to distinguish the two observed regimes
of quantum turbulence: Kolmogorov turbulence and Vinen turbulence.
Some of the tools are not standard, for example smoothed vorticity, or
the computational use of crossing numbers in dense configurations of field
lines. Some tools, for example the use of knot polynomials, need to
be applied more systematically to a much wider range of systems to make useful
comparisons and learn which is the most useful. 

The most urgent exploration would be to check if some of the tools 
presented here
can be applied to classical turbulence,
perhaps only in a statistical sense, since in a classical fluid
vortex lines may not be traceable
as robustly as in a quantum fluid.

\begin{acknowledgement}
I am grateful to collaborators who taught me many things, 
particularly David Samuels, Andrew Baggaley, Luca Galantucci and Ladik Skrbek.
The work of Mae Mesgarnezhad and Rob Cooper was essential for 
Section \ref{sec:topology}.
 Above all, I am indebted to Renzo Ricca, whose enthusiasm 
and focus on what is important triggered and sustained my interest
in topological fluid dynamics. The support of UKRI is acknowledged.
\end{acknowledgement}

\newpage

%
%

\begin{thebibliography}{99.}


\bibitem{GlatzmaierRoberts}
Glatzmaier, G.A., and Roberts, P.H.:
A three-dimensional self-consistent computer simulation of a
geomagnetic field reversal.
Nature, {\bf 377},  203 (1995).


\bibitem{Pitaevskii}
Pitevskii, L., and Stringari, S.: {\it Bose-Einstein condensation}.
Clarendon Press, Oxford (2003).

\bibitem{primer}
Barenghi, C.F., and Parker, N.G.: {\it A primer on quantum fluids}.
Springer (2016).


\bibitem{Vinen1961}
Vinen, W.F.:
The detection of single quanta of circulation in liquid helium~II.
Proc. Roy. Soc. A {\bf 260}, 218 (1961).

\bibitem{Hough2001}
Hough, L., Donev, L.A.K., and Zieve, R.J.:
Smooth vortex precession in superfluid $^4$He,
Phys. Rev. B {\bf 65}, 024511 (2001).

\bibitem{Schwab1997}
Schwab, K., Bruckner, N., and  Packard, R.E.:
Detection of the Earth's rotation using superfluid phase coherence.
Nature {\bf 386}, 585 (1997).

\bibitem{Pomeau}
Pomeau, Y., and Rica, S.:
Transition to dissipation in a model of superflow.
Phys. Rev. Lett. {\bf 69}, 1644 (1992).

\bibitem{Keepfer}
Keepfer, N.A., Stagg, G.W., Galantucci, L., 
Barenghi, C.F., and  Parker, N.G.:
Spin-up of a superfluid vortex lattice driven by rough boundaries.
Phys. Rev. B {\bf 102}, 144520 (2020)

\bibitem{Comaron2019}
Comaron, P., Larcher, F., Dalfovo, F., and  Proukakis, N.P.:
Quench dynamics of an ultracold two-dimensional Bose gas.
Phys. Rev. A {\bf 100}, 033618 (2019).

\bibitem{DonnellyDonnelly1981}
Donnelly, R.J.,  Donnelly, J.A.,  and  Hills, R.N.:
Specific heat and dispersion curve for helium II.
J. Low Temp. Phys. {\bf 44}, 471 (1981).

\bibitem{GalliReattoRossi2014}
Galli, D.E., Reatto, L., and Rossi, M.:
Quantum Monte Carlo study of a vortex in superfluid $^4$He and
search for a vortex state in the solid.
Phys. Rev. B {\bf 89}, 224516 (2014).

\bibitem{Amelio2018}
Amelio, I.,  Galli, D.E., and Reatto, L.:
Probing quantum turbulence in $^4$He by quantum evaporation measurements.
Phys. Rev. Lett. {\bf 121}, 015302 (2018)

\bibitem{BerloffRoberts1999}
Berloff, N.G., and Roberts, P.H.:
Motions in a Bose condensate: VI. Vortices in a nonlocal model.
J. Phys. A. Math Gen. {\bf 32}, 5611 (1999).

\bibitem{OrtizCeperley1995}
Ortiz, G. and Ceperley, D.M.:
Core structure of a vortex in superfluid $^4$He.
Phys. Rev. Lett. {\bf 75}, 4642 (1995).

\bibitem{SaddChesterReatto1997}
Sadd, M., Chester, G.V., and Reatto, L.:
Structure of a vortex in superfluid $^4$He.
Phys. Rev. Lett. {\bf 79}, 2490 (1997).

\bibitem{friction}
Barenghi, C.F.,  Vinen, W.F., and Donnelly, R.J.:
Friction on quantized vortices in Helium II: a review,
J. Low Temp. Phys. {\bf 52}, 189-247 (1982).

\bibitem{White2010}
White, A., Proukakis. N.P., and  Barenghi, C.F.;
Non classical velocity statistics in a turbulent atomic
Bose Einstein condensate,
Phys. Rev. Lett. {\bf 104}, 075301 (2010).

\bibitem{Okamoto}
Okamoto, N., Yoshimatsu, K., Schneider, K., Farge, M., and Kaneda, Y.:
Coherent vortices in high resolution direct numerical simulation 
of homogeneous isotropic turbulence: a wavelet viewpoint.
Phys. Fluids {\bf 19}, 115109 (2007).

\bibitem{Frisch}
Frisch, U.: {\it Turbulence. The legacy of A.N. Kolmogorov},
Cambridge University Press (1995).

\bibitem{Tabeling}
Maurer, J., and Tabeling, P.:
Local investigation of superfluid turbulence.
Europhys. Lett. {\bf 43}. 29 (1998).

\bibitem{Salort}
Salort,J., Chabaud,B., L\'{e}v\^{e}que, E., and  Roche, P.-E.:
Energy cascade and the four-fifths law in superfluid turbulence.
Europhys. Lett. {\bf 97}, 34006 (2012).


\bibitem{Bradley}
Bradley, D.I., Fisher, S.N.,  Gu\'{e}nault, A.M.,  Haley, R.P.,
Pickett, G.R., Potts, D., and Tsepelin, V.
Direct measurement of the energy dissipated by quantum turbulence.
Nature Phys. {\bf 7}, 473 (2011).

\bibitem{Nore}
Nore, C.,. Abid, M., and Brachet, M.E.:
Kolmogorov turbulence in low-temperature superflows
Phys. Rev. Lett.{\bf 78}, 3896 (1997).

\bibitem{Araki}
Araki, T., Tsubota, M., and Nemirovskii, S.K.:
Energy spectrum of superfluid turbulence with no normal-fluid component.
Phys. Rev. Lett. {\bf 89}, 145301 (2002).

\bibitem{Kobayashi}
Kobayashi, M., and Tsubota, M.:
Quantum turbulence in a trapped Bose-Einstein condensate.
Phys. Rev. A {\bf 76}, 045603 (2007).

\bibitem{Baggaley-EPL}
Baggaley, A.W., Barenghi, C.F., Shukurov, A., and Sergeev, Y.A.:
Coherent vortex structures in quantum turbulence
Europhys. Lett. {\bf 98}, 26002 (2012).

\bibitem{BarenghiLvovRoche}
Barenghi, C.F., L'vov, V., and  Roche, P.-E.:
Experimental, numerical and analytical velocity spectra in
turbulent quantum fluid.
Proc. Nat. Acad. Sci. USA, {\bf 111} (Suppl. 1) 4683 (2014).

\bibitem{Baggaley2014}
Baggaley, A.W.: The importance of vortex bundles in quantum turbulence
at absolute zero.
Phys. Fluids {\bf 24}, 055109 (2012).

\bibitem{Laurie}
Andrew W. Baggaley, A.W., Laurie, J., and Barenghi, C.F.:
Vortex-density fluctuations, energy spectra, and vortical regions 
in superfluid turbulence.
Phys. Rev. Lett. {\bf 109}, 205304 (2012).

\bibitem{Roche}
Roche, P.-E., and Barenghi, C.F.:
Vortex spectrum in superfluid turbulence: interpretation of
a recent experiment.
Europhys Lett. {\bf 81}, 36002, (2008).

\bibitem{Paoletti}
Paoletti, M.S., Fisher, M.E., Sreenivasan, K.R., and Lathrop, D.P.:
Velocity statistics distinguish quantum turbulence from classical turbulence,
Phys. Rev. Lett. 101, 154501 (2008).

\bibitem{Vincent}
Vincent, A., and Meneguzzi, M.:
The spatial structure and statistical properties of homogeneous turbulence.
J. Fluids Mech. {\bf 225}, 1 (1991).

\bibitem{stats}
Baggaley, A.W., and Barenghi, C.F.:
Quantum turbulent velocity statistics and quasiclassical limit.
Phys. Rev. E {\bf 84}, 067301 (2011).

\bibitem{LaMantia}
La Mantia, M., and Skrbek, L.:
Quantum, or classical turbulence? 
Europhys. Lett. {\bf 102}, 46002, (2014).

\bibitem{Skrbek2021}
Skrbek, L., Schmoranzer, D., Midlik, S., and  Sreenivasan, K.R.:
Phenomenology of quantum turbulence in superfluid helium.
Proc. Nat. Acad. Sci. USA {\bf 118}, 16 (2021).

\bibitem{Farge}
Farge, M., Pellegrino, G., and Schneider, K.:
Coherent vortex extraction in 3D turbulent flows using orthogonal wavelets.
Phys. Rev. Lett. {\bf 87}, 054501 (2001).

\bibitem{Volovik}
Volovik, G.E.:
On developed superfluid turbulence.
J. Low Temp. Phys. {\bf 136}, 309 (2004).

\bibitem{Walmsley}
Walmsley, P.M., and Golov, A.I.:
Quantum and quasiclassical types of superfluid turbulence.
Phys. Rev. Lett. {\bf 100}, 245301 (1008).

\bibitem{Bradley}
Bradley, D.I., Clubb, D.O., Fisher, S.N.,
Gu\'{e}nault, A.M., Haley, R.P., Matthews, C.J.,
Pickett, G.R., Tsepelin, V., and Zaki, K.:
Decay of pure quantum turbulence in superfluid $^3$He-B.
Phys. Rev. Lett. {\bf 96}, 035301 (2006).

\bibitem{BaggaleyPRB2012}
Baggaley, A.W.,  Barenghi, C.F., and  Sergeev, Y.A.:
Quasiclassical and ultraquantum decay of superfluid turbulence.
Phys. Rev. B {\bf 85}, 060501(R) (2012).

\bibitem{Stagg}
Stagg, G.W.,  Parker, N.G., and  Barenghi, C.F.:
Ultraquantum turbulence in a quenched homogeneous Bose gas
Phys. Rev. A {\bf 94}, 053632 (2016).

\bibitem{Cidrim}
Cidrim, A., White, A.C.,  Allen, A.J.,
Bagnato, V.S., and  Barenghi, C.F.:
Vinen turbulence via the decay of multi-charged vortices in trapped
Bose-Einstein condensates.
Phys. Rev. A 96, 023617 (2017).


\bibitem{Mocz}
Philip Mocz, P., Vogelsberger, M., Robles, V.H.,
Zavala, J., Boylan-Kolchin, M., Fialkov, A., and Hernquist, L.:
Galaxy formation with BECDM – I. Turbulence and relaxation of
idealized haloes.
M.N.R.A.S {\bf 471}, 4559 (2017).

\bibitem{nocascade}
Barenghi, C.F., Sergeev,Y.A., and Baggaley, A.W.:
Regimes of turbulence without an energy cascade.
Sci. Reports {\bf 6}, 35701 (2016).

\bibitem{Hanninen}
Barenghi, C.F., H\"anninen, R., and Tsubota, M.:
Anomalous translational velocity of vortex ring with finite-amplitude
Kelvin waves.
Phys. Rev. E {\bf 74}, 046303 (2006).

\bibitem{Vinen2001}
Vinen, W.F.:
Decay of superfluid turbulence at a very low temperature: 
the radiation of sound from a Kelvin wave on a quantized vortex.
Phys. Rev. B {\bf 64}, 134520 (2001).

\bibitem{Parker}
Barenghi, C.F., Parker, N.G., Proukakis, N.P.,
and Adams, C.S.:
Decay of quantised vorticity by sound emission.
J. Low Temp. Phys. {\bf 138}, 629 (2005).

\bibitem{Svistunov}
Svistunov, B.V.:
Superfluid turbulence in the low-temperature limit.
Phys. Rev. B {\bf 52}, 3647 (1995).

\bibitem{Leadbeater}
Leadbeater, M., Winiecki, T.,  Samuels, D.C.,
Barenghi, C.F., and Adams, C.S.:
Sound emission due to superfluid vortex reconnections.
Phys. Rev. Lett. {\bf 86} 1410 (2001).

\bibitem{Zuccher}
Zuccher, S., Caliari, M.,  Baggaley. A.W., and  Barenghi, C.F.:
Quantum vortex reconnections,
Phys. Fluids {\bf 24}, 125108 (2012).

\bibitem{reconnections}
Galantucci, L.,  Baggaley, A.W., Parker, N.G., and  Barenghi, C.F.:
Crossover from interaction to driven regimes in quantum vortex reconnections.
Proc. Nat. Acad. Sci. USA {\bf 116}, 12204 (2019).

\bibitem{Morishita}
Morishita, M., Kuroda, T., Sawada, A., and Satoh, T.:
Mean free path effects in superfluid $^4$He,
J. Low Temp. Phys. {\bf 76}, 387 (1989).

\bibitem{Schwarz}
Schwarz, K.W.:
Three-dimensional vortex dynamics in superfluid $^4$He: 
homogeneous isotropic turbulence. Phys. Rev. B {\bf 38}, 2398 (1988).

\bibitem{Kida}
Kida, S.: A vortex filament moving without change of form.
J. Fluid Mech. {\bf 112}, 397 (1985).

\bibitem{Ricca-Samuels-Barenghi}
Ricca, R.L., Samuels, D.C., and Barenghi, C.F.:
Evolution of vortex knots.
J. Fluid Mech. {\bf 391}, 29 (1999).

\bibitem{Proment}
Proment, D., Onorato, M., and  Barenghi, C.F.:
Vortex knots in a Bose-Einstein
condensate.
Phys. Rev. E {\bf 85}, 036306 (2012).

\bibitem{Oberti}
Oberti, C., and Ricca, R.L.:
Influence of winding number on vortex knots dynamics.
Sci. Reports {\bf 9}, 1 (2019).

\bibitem{Kleckner-untie}
Kleckner, D., Kauffman, L.H., and  Irvine, W.T.M.:
How superfluid vortex knots untie,
Nat. Phys. {\bf 12}, 650 (2016).

\bibitem{LiuRicca2016}
Liu, X. and Ricca, R.L.:
Knots cascade detected by a monotonically decreasing sequence of values.
Sci. Rep. {\bf 6}, 24118 (2016).

\bibitem{LiuRiccaLi}
Xin Liu, X., Ricca, R.L., and Li, X.-F.:
Minimal unlinking pathways as geodesics in knot polynomial space.
Comm. Phys. {\bf 3}, 1 (2020)

\bibitem{Kleckner-knot}
Kleckner, D. and  Irvine, W.T.M.:
Creation anddynamics of knotted vortices.
Nat. Phys. {9}, 253 (2013).

\bibitem{Mae}
Mesgarnezhad, M., Cooper, R.G., Baggaley, A.W.,
and Barenghi, C.F.:
Helicity and topology of a small region of quantum vorticity.
Fluid Dyn. Res. {\bf 50}, 011403 (2018).


\bibitem{Cooper}
Cooper, R.G., Mesgarnezhad, M.,
Baggaley, A.W., and Barenghi, C.F.:
Knot spectrum of turbulence.
Sci. Reports {\bf 9}, 10545 (2019)

\bibitem{Barenghi-Ricca-Samuels}
Barenghi, C.F., Ricca, R.L., and  Samuels, D.C.:
How tangled is a tangle?
Physica D {\bf 157}, 197 (2001).

\bibitem{Moffatt1969}
Moffatt, H.K.: Degree of knottedness of tangled vortex line,
J. Fluid Mech. {\bf 35}, 117 (1969).

\bibitem{diLeoni2017}
Clark di Leoni, P., Mininni, P.D., and  Brachet, M.E.:
Dual cascade and dissipation mechanisms in helical quantum turbulence.
Phys. Rev. A {\bf 95}, 053636 (2017).

\bibitem{MoffattRicca1992}
Moffatt, H.K., and Ricca, R.L.:
Helicity and the C\u{a}lug\u{a}reanu invariant,
Proc. Roy. Soc. London A {\bf 439},  411 (1992).

\bibitem{Hanninen2016}
H\"{a}nninen, R., Hietala, N., and Salman, H.:
Helicity within the vortex filament model.
Sci. Reports {\bf 6}, 37571 (2016).

\bibitem{ZuccherRicca2018}
Zuccher, S. and  Ricca, R.L.:
Twist effects in quantum vortices and phase defects.
Fluid Dyn. Res. {\bf 50}, 011414 (2018).

\bibitem{Salman2018}
Salman,  H.:
Helicity conservation and twisted Seifert surfaces for superfluid
vortices.
Proc. Roy. Soc. A {\bf 473}, 20160853 (2018)

\bibitem{Galantucci2021}
Galantucci, L.,  Barenghi, C.F.,  Parker, N.G.,
and  Baggaley, A.W.:
Mesoscale helicity distinguishes Vinen from Kolmogorov turbulence in helium-II.
Phys. Rev. B {\bf 103}, 144503 (2021).

\bibitem{Livingstone}
Livingstone, L.: {\it Knot Theory} Cambridge University Press, 
Cambridge (1993).

\bibitem{Arsuaga}
Arsuaga, J., Vasquez, M., Trigueros, S.,
Summers, D.W., and Roca, J.:
Knotting probability of DNA molecules confined in restricted volumes: DNA knotting in phage capsds. Proc. Nat. Acad. Sci. USA {\bf 99}, 5373 (2002).

\bibitem{Raymer}
Raymer, D.M, and Smith, M.:
Spontaneous knotting of an agitated string.
Proc. Nat. Sci. Acad. Sci. USA {\bf 104}, 16432 (2007).

\end{thebibliography}
%

\end{document}